\documentclass[twocolumn]{aastex631}

\usepackage{graphicx}
\usepackage[english]{babel}
\usepackage{subfigure}
\usepackage{amsmath} 
\usepackage{natbib}
\usepackage{xspace}

\newcommand{\IXPE}{\textit{IXPE}\xspace}

\newcommand{\GRS}{\textit{GRS}\xspace}
\newcommand{\MAXI}{\textit{MAXI}\xspace}
\newcommand{\NICER}{\textit{NICER}\xspace}

\newcommand{\RN}[1]{%
  \textup{\uppercase\expandafter{\romannumeral#1}}%
}

\shorttitle{Double Lamp post}
\shortauthors{Lucchini et al.}

\usepackage{lipsum}  

\graphicspath{{./}{figures/}}
\begin{document}

\title{Investigating the impact of vertically extended coronae on X-ray reverberation mapping}

%idk what the number under here is, probably the orcid
\author[0000-0002-2235-3347]{Matteo Lucchini}
\affiliation{MIT Kavli Institute for Astrophysics and Space Research, MIT, 70 Vassar Street, Cambridge, MA 02139, USA}

\author[0000-0003-4216-7936]{G~Mastroserio}
\affiliation{Cahill Center for Astronomy and Astrophysics, California Institute of Technology, Pasadena, CA 91125, USA}
\affiliation{INAF-Osservatorio Astronomico di Cagliari, via della Scienza 5, I-09047 Selargius (CA), Italy}

\author{Jingyi Wang}
\affiliation{MIT Kavli Institute for Astrophysics and Space Research, MIT, 70 Vassar Street, Cambridge, MA 02139, USA}

\author{Erin Kara}
\affiliation{MIT Kavli Institute for Astrophysics and Space Research, MIT, 70 Vassar Street, Cambridge, MA 02139, USA}

\author{Adam Ingram}
\affiliation{Department of Physics, Astrophysics, University of Oxford, Denys Wilkinson Building, Keble Road, Oxford OX1 3RH, UK}
\affiliation{School of Mathematics, Statistics and Physics, Newcastle University, Herschel Building, Newcastle upon Tyne, NE1 7RU, UK}

\author{Javier Garcia}
\affiliation{Cahill Center for Astronomy and Astrophysics, California Institute of Technology, Pasadena, CA 91125, USA}

\author{Thomas Dauser}
\affiliation{Dr. Karl Remeis-Observatory and Erlangen Centre for Astroparticle Physics, Friedrich-Alexander-Universit\"at Erlangen-N\"urnberg, Sternwartstr.~7, 96049 Bamberg, Germany}

\author{Michiel van der Klis}
\affiliation{Anton Pannekoek Institute for Astronomy, University of Amsterdam, Science Park 904, NL-1098 XH Amsterdam, Netherlands}

\author{Ole K\"{o}nig}
\affiliation{Dr. Karl Remeis-Observatory and Erlangen Centre for Astroparticle Physics, Friedrich-Alexander-Universit\"at Erlangen-N\"urnberg, Sternwartstr.~7, 96049 Bamberg, Germany}

\author{Collin Lewin}
\affiliation{MIT Kavli Institute for Astrophysics and Space Research, MIT, 70 Vassar Street, Cambridge, MA 02139, USA}

\author{Edward Nathan}
\affiliation{Department of Physics, Astrophysics, University of Oxford, Denys Wilkinson Building, Keble Road, Oxford OX1 3RH, UK}
\affiliation{Cahill Center for Astronomy and Astrophysics, California Institute of Technology, Pasadena, CA 91125, USA}

\author{Christos Panagiotou}
\affiliation{MIT Kavli Institute for Astrophysics and Space Research, MIT, 70 Vassar Street, Cambridge, MA 02139, USA}

\begin{abstract}

Accreting black holes commonly exhibit hard X-ray emission, originating from a region of hot plasma near the central engine referred to as the corona. The origin and geometry of the corona are poorly understood, and models invoking either inflowing or outflowing material (or both) can successfully explain only parts of the observed phenomenology. In particular, recent works indicate that the time-averaged and variability property might originate in different regions of the corona. In this paper we present a model designed to move beyond the lamp post paradigm, with the goal of accounting for the vertical extent of the corona. In particular, we highlight the impact of including self consistently a second lamp post, mimicking for example an extended jet base. We fully include the effect that the second source has on the time-dependent disk ionization, reflection spectrum, and reverberation lags. We also present an application of this new model to NICER observations of the X-ray binary MAXI J1820+070 near its hard-to-soft state transition. We demonstrate that in these observations, a vertically extended corona can capture both spectral and timing properties, while a single lamp post model can not. In this scenario, the illumination responsible for the time-averaged spectrum originates close to the black hole, while the variability is likely associated with the ballistic jet. 
\end{abstract}

\keywords{accretion --- black hole physics --- reverberation mapping}

\section{Introduction} 
\label{sec:intro}

Accreting black holes commonly display luminous X-ray emission. In Black Hole X-ray Binaries (BHBs) and Active Galactic Nuclei (AGN), the hard ($\geq$ a few keV) X-ray emission is typically generated in a region referred to as the corona: a population of hot electrons ($kT_{\rm e}\approx100$ keV) located in a compact, optically thin ($\tau\approx1$) region \citep[e.g][]{Eardley75}. These electrons inverse-Compton scatter soft seed photons likely emitted by an optically thick, geometrically thin accretion disk \citep{Shakura73} into a power-law spectrum extending from a few keV, up to tens to hundreds of keV. This non-thermal emission is typically referred to as the continuum emission.

Beyond this simple picture, the nature of the corona and the details of the physics powering it remain poorly understood. The three most common models postulate that the corona is either a slab-like atmosphere which sandwiches the disk \citep[e.g.][]{Haardt93,Haardt94}, a hot, geometrically thick, optically thin accretion flow \citep[e.g][]{Narayan96,Esin97}, or a compact source located on the spin axis of the black hole (typically represented by a point source: the lamp post model, e.g. \citealt{Matt91,Beloborodov99}) at a height $h$, which might be related the base of the jet \citep{Markoff05}. Regardless of the origin of the corona, some of the continuum photons, rather than reaching the observer directly, hit the disk, where they are reprocessed. The result of this reprocessing is commonly referred to as the reflection spectrum, which is made up of three main components: the Compton hump, a broad peak around 20--30 keV; the $\sim 6.4-6.9$ keV iron K-$\alpha$ line complex, and a multitude of other spectral lines below 1 keV \citep[e.g.][]{Fabian89,Garcia10}. The coronal emission typically originates close to the black hole, and therefore so does the reflection, as relativistic effects bend the photon trajectories towards the innermost regions of the disk. Gravitational red-shifts and relativistic velocities smear the emission lines; as a result, the iron line is observed as a broad feature, and at low energy the various other lines are blurred together in a smooth continuum \citep{Laor91,Fabian00,Garcia16}. The observed reflection signal therefore carries information about the innermost regions of the accretion flow, as well as of the geometry of the space-time around the black hole \citep{Wilms01,Reynolds08,Dauser13,Garcia15,Bambi17}. 

The coronal continuum is highly variable, which also causes the reflection to vary. Crucially, the variability of the latter, with respect to the former, is delayed (to first order) by the light-travel time from the corona to the disk. These delays are referred to as reverberation lags, and offer additional information on the geometry of the accreting material \citep{McHardy07,Fabian09,Uttley11,DeMarco17,Kara19}. Reverberation lags, however, are not easily identified in the time domain. This is because the light curve also includes large amounts of variability intrinsic to the corona, which drives additional lags likely due to the propagation of fluctuations in the mass accretion rate through the system \citep{Lyubarskii97,Kotov01,Arevalo06,Rapisarda14}. The two types of lags are typically disentangled through Fourier analysis, as the reverberation signal dominates at high Fourier frequencies, while it is diluted by the coronal variability at low Fourier frequencies \citep[e.g.][]{Nowak99,Pottschmidt00,McHardy04,DeMarco17,Kara19,Wang22}. Performing spectral-timing analysis by modeling both the time-averaged and time-dependent signals, while disentangling the continuum lags from the reverberation lags, is therefore a powerful tool to fully map the accretion flow, as well as measure physical parameters of the system like the black hole mass and spin \citep{Chainakun16,CaballeroGarcia18,Mastroserio19,Alston20,Ingram22}. However, applying this approach with current reflection models is not straightforward. While for some sources it is possible to reproduce both the timing and spectral data (e.g. in Cygnus X-1, \citealt{Mastroserio19}, or Ark 564, \citealt{Lewin22}), in a few sources the coronal size/height inferred from the reverberation lags is much larger than that from the time-averaged spectra \citep[e.g.][]{Zoghbi21,Wang21}. 

The main issue faced by current spectral-timing reflection models is in the simplified treatment of the corona, particularly its geometry. Notably, the lamp post geometry is practical because the calculation of the relativistic effects is greatly simplified: the height $h$ effectively acts as a way of setting the reflection emissivity profile, with lower heights corresponding to steeper emissivities (meaning that more of the reflection originates in the innermost regions of the disk, e.g. \citealt{Dauser13,Dauser22}). Nevertheless, it is clear that treating the corona as a point source is an over-simplification, and indeed various models for extended coronae (discussed below) have been presented in the literature. 

The most complete model presented to date is arguably that of \cite{Wilkins16}, who account for propagating fluctuations in a horizontally and vertically extended corona, as well as the reverberation signal driven from it. While the model is too computationally expensive to be easily fit to data, the authors concluded that both the radial and vertical extent of the corona play an important role in shaping the lag spectra. \cite{Chainakun17} consider only the vertical extent of the corona, using a two-blob approximation similar to that treated here. By fitting individual lag-frequency or lag-energy spectra of the Seyfert galaxy PG 1244$+$026, these authors find that the corona is likely vertically extended, similarly to the base of a jet. Crucially, due to the computational cost of the model, these authors do not fit time-averaged and lag spectra simultaneously. Instead, they ensure a posteriori that the parameters inferred from fitting the timing data are consistent with the time-averaged spectra. The main differences between our work and theirs is that our model includes additional physical processes responsible for driving lags (detailed in Sec.\ref{sec:model}) as well as high disk densities. Additionally, we fit our model to multiple time-averaged and lag-energy spectra simultaneously. Other mathematically similar models (which however treat the reflection component in less detail) split the corona into two radial zones instead of two vertical zones \citep{Kawamura22,Kawamura23}.

\begin{figure}
    \centering
    \includegraphics[width=\columnwidth, trim={0.0 0.0cm 0.0cm 1.5cm},clip]{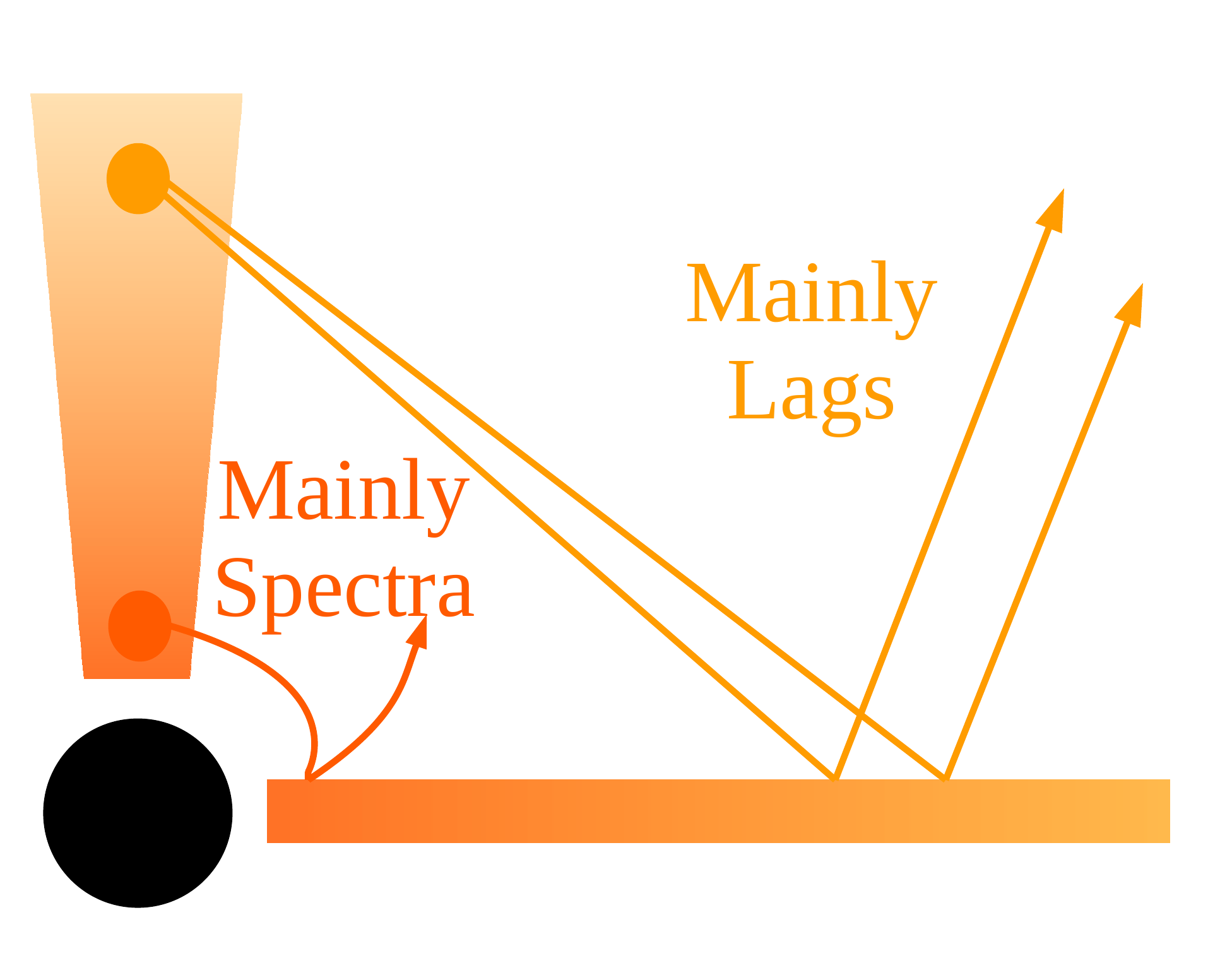}
    \caption{Schematic of the model presented in this work, and inferred from the recent \NICER observations of multiple X-ray binaries \protect\citep{Wang21,Wang22}.}
    \label{fig:schematic}
\end{figure}

Recently, \cite{Bellavita22} extended the variable Comptonization model of \cite{Karpouzas20} and applied it to the type-B Quasi Periodic Oscillation (QPO) of \MAXI J1348$-$630 as well as the type-C QPO of \GRS 1915$+$105, finding that in both cases a two-zone, vertically extended corona is favoured by the data. A vertically extended corona also appears to be required in order to explain the behavior of the reverberation lags shown by BHB \MAXI J1820$+$070 \citep{Kara19,Wang21,DeMarco21}, as well as the full \NICER sample of sources with detected reverberation lags \citep{Wang22} (although we note that the first \IXPE observations of Cyg X-1 seem to favor a horizontally extended corona, \citealt{Ixpe1}). In order to reconcile the steep emissivity profile inferred from time-averaged data with the long reverberation delays inferred from the lags, in this work we test a scenario in which the time-averaged reflection is dominated by a region located a few $R_{\rm g}=GM/c^{2}$ (where $G$ is the gravitational constant, $M$ the mass of the black hole and $c$ the speed of light) away from the black hole (possibly where the jet is launched in the first place), while the reverberation lags originate further downstream in the outflow. In our model, the long lags originating from this larger region dominate over the short lags coming from near the black hole, where most of the time-averaged reflection spectrum is produced by photons focused towards the inner disk by GR effects. Our model is illustrated in Fig.\ref{fig:schematic}.

Here we present an initial proof of concept for accounting for the vertical extension of the corona in the public reverberation model \texttt{reltrans} \citep{Ingram19}, by self-consistently including the signal driven by a second lamp post. The new model flavor discussed here is called \texttt{rtransDbl}. The paper is structured as follows: in section \ref{sec:model} we describe the additional formalism introduced by the inclusion of a second X-ray source, in section \ref{sec:behavior} we discuss the behavior of the new model and compare it to the single lamp post scenario, in section \ref{sec:1820} we apply our new model to one \NICER observation of \MAXI J1820$+$070 taken during the hard-to-soft state transition, in section \ref{sec:discussion} we discuss our findings, and in section \ref{sec:conclusion} we draw our conclusions.

\section{Model description}
\label{sec:model}

In \texttt{reltrans}, the variability of the continuum is quantified through a pivoting power-law, in which both the coronal normalization and photon index are allowed to vary through a simple phenomenological parametrization \citep{Kording04,Ingram19}. The disk response to these fluctuations, which includes the light travel time differences, as well as the changes in the reflection spectrum driven by the changes in the continuum, are then computed self consistently through a transfer function \citep{Campana95,Reynolds99,Wilkins13}. The rest-frame reflection spectrum from the disk is computed by using the \texttt{xillver} reflection model \citep{Garcia10}, and relativistic effects are calculated on the fly using the \texttt{YNOGK} code \citep{Yang13}. The combination of a more physical coronal geometry with the low computational cost of \texttt{reltrans} allows us to thoroughly test the scenario described above by jointly fitting time-averaged spectra with lag-energy spectra taken at multiple Fourier frequencies.

\subsection{Continuum}
\label{sec:cont}

In this section we describe the extension of the pivoting power-law continuum assumed in \texttt{reltrans}, in order to include a second point source of X-rays. We use the \texttt{nthcomp} model \citep{Zdziarski96} to calculate the continuum spectral shape in our code, but use a cutoff powerlaw in the equations throughout the paper for clarity. However, the derivations presented in this section are generic. In this first iteration of the model we assume that the two sources are entirely incoherent from each other, meaning that they are allowed vary independently. This assumption means that the variability is intrinsic to each source. This scenario could be realized, for example, if the heating/cooling of each region in the corona is driven locally by magnetic reconnection or turbulence \citep{Beloborodov17}. Finally, we allow each source to contribute differently to the time-averaged and time-dependent signals. This choice allows us to explore scenarios in which part of the corona is more variable but not significantly brighter than the other, thus dominating the lags while only having a minor effect on the time-averaged spectra. We note that our formalism can in principle be generalized to an arbitrary number $n$ of lamp posts, although the model run time (and number of parameters) will increase significantly. The specific energy flux of the continuum seen by an observer on Earth can be written as:
\begin{align}
    F(E,t) = C_1(t) l_1 g_{\rm s_1 o}^{\Gamma(t)} E^{1-\Gamma(t)} e^{-E/E_{\rm{cut,obs,1}}} + \nonumber \\
    C_2(t-\tau_{d}) l_2 g_{\rm s_2 o}^{\Gamma(t-\tau_{d})} E^{1-\Gamma(t-\tau_{d})} e^{-E/E_{\rm{cut,obs,2}}};
    \label{eq:cont_full}
\end{align}
eq.\ref{eq:cont_full} is evaluated in the observer frame, and therefore  the time $\tau_{d}=\tau_{\rm s_2 o} -\tau_{\rm s_1 o}$ must be included to account for the difference between photon arrival times from the two sources, and $\tau_{\rm s_j o}$ is the time the photons take to travel from the $j$-th source to the observer. $E_{\rm cut,obs,j} = g_{\rm s_j o} E_{\rm cut}$ is the cutoff in the spectrum of each lamp post, observed in the lab frame, and $E_{\rm cut}$ is the cutoff in each source frame. $l_j$ and $g_{\rm s_j o}$ are the lensing factors and gravitational redshifts, and for a given source depend exclusively on the source height $h_j$ of the $j$-th lamp post. $C_j(t)$ is the normalization of each continuum spectrum, and $\Gamma(t)$ the photon index. As a convention, we take the subscript $2$ to indicate the higher lamp post and $1$ to indicate the lower one throughout the paper. We use the subscript $0$ to indicate time-averaged quantities. Finally, for simplicity we assume that both lamp posts have identical photon index $\Gamma$ and cutoff energy $E_{\rm cut}$ (in their co-moving frame, meaning that the cutoff of each spectrum depends on $h_j$). To simplify the notation in the rest of this section, we define the energy-dependence for the continuum of each lamp post $j$ as:
\begin{equation}
    D_j(E) = l_j g_{\rm s_j o}^{\Gamma_0} E^{1-\Gamma_0}e^{-E/E_{\rm cut,obs,j}},
    \label{eq:cont_shape}
\end{equation}
where $\Gamma_0$ is the average photon index. The time-dependent continuum spectrum can be re-written as:
\begin{align}
    F(E,t) = C_1(t) D_1(E) g_{\rm s_1 o}^{\delta \Gamma(t)} E^{-\delta\Gamma(t)} + \nonumber \\
    C_2(t-\tau_{d}) D_2(E) g_{\rm s_2 o}^{\delta \Gamma(t-\tau_{d})} E^{-\delta\Gamma(t-\tau_{d})},
    \label{eq:cont_delta}
\end{align}
where $\delta\Gamma(t) = \Gamma(t)-\Gamma_0$. Assuming the variation in $\Gamma$ is small, we can Taylor expand Eq.~\ref{eq:cont_delta} to first order, around $\Gamma_0$:
\begin{align}
   F(E,t) = C_1(t)D_1(E)\left[1-\ln{\left(\frac{E}{g_{\rm s_1 o}}\right)}\delta\Gamma(t)\right] + \nonumber \\
   C_2(t-\tau_{d})D_2(E)\left[1-\ln{\left(\frac{E}{g_{\rm s_2 o}}\right)}\delta\Gamma(t-\tau_{d})\right]
   \label{eq:cont_ln}
\end{align}
Defining for each lamp post $j$: $C_j(t) = C_{j,0} + \delta C_j(t)$ and only considering first order terms, Eq.~\ref{eq:cont_ln} becomes:
\begin{align}
    F(E,t) =  C_{1,0}D_1(E)\left[1+\frac{\delta C_1(t)}{C_{1,0}} - \ln{\left(\frac{E}{g_{\rm s_1 o}}\right)}\delta\Gamma(t)\right] + \nonumber \\
    C_{2,0}D_2(E)\left[1+\frac{\delta C_2(t-\tau_{d})}{C_{2,0}} - \ln{\left(\frac{E}{g_{\rm s_2 o}}\right)}\delta\Gamma(t-\tau_{d})\right]
    \label{eq:cont_taylor}
\end{align}
The Fourier transform of Eq.~\ref{eq:cont_taylor} (for $\nu > 0$) is:
\begin{align}
    F(E,\nu) = D_1(E)\left[ C_1(\nu) - C_{1,0} \ln{\left(\frac{E}{g_{\rm s_1 0}}\right)}\Gamma(\nu) \right] + \nonumber \\
    e^{i2\pi\tau_{d}\nu} D_2(E)\left[ C_2(\nu) - C_{2,0} \ln{\left(\frac{E}{g_{\rm s_2 0}}\right)}\Gamma(\nu) \right].
\end{align}
Finally, we characterize the pivoting of the two continuum spectra by defining $\Gamma(\nu) C_{1,0}/C_1(\nu) = \gamma_1(\nu) e^{i \phi_{\rm AB,1}(\nu)}$ and $\Gamma(\nu) C_{2,0}/C_2(\nu) = \gamma_2(\nu) e^{i \phi_{\rm AB,2}(\nu)}$. These parameters have identical meaning to older model flavors. $\gamma_j(\nu)$ quantifies the fractional variability of the continuum photon index from each source, with respect to the fractional variability of the continuum normalization, and $\phi_{\rm AB,j}(\nu)$ is the phase difference between variations in the photon index and normalization of each coronal spectrum \citep{Mastroserio21}. Finally, we define the normalization ratio parameter $\eta(\nu) = C_{2}(\nu)/C_{1}(\nu)$, which quantifies how much of the total continuum variability in a given Fourier frequency range is driven by each lamp post. For $\nu=0$ (the time-averaged spectrum), $\eta(\nu=0)\equiv\eta_{0}$ simply sets the ratio of the average normalizations of each X-ray source; $\eta_0=0$ recovers the continuum from the lower lamppost, and $\eta=\infty$ that from the upper lamppost. As a result, $\eta_0$ also sets how much each of the two sources contributes to the time-averaged reflection spectrum, as well as the disk ionization and cutoff energy observed in the disk frame, as we detail in the following section. On the other hand, in the timing domain $\eta(\nu)$ sets how much the signal from each source contributes to the lags. Neglecting reflection, the total continuum contribution in the Fourier domain is:
\begin{align}
    F(E,\nu) = C_1(\nu)\bigg[D_1(E) + e^{i2\pi\tau_{d}\nu} \eta(\nu) D_2(E) -  \nonumber \\
     \gamma_{1}(\nu)e^{i\phi_{\rm AB_{1}}(\nu)} D_1(E)\ln{\left(\frac{E}{g_{\rm s_1 0}}\right)} -  \nonumber \\
  e^{i2\pi\tau_{d}\nu} \gamma_{2}(\nu)e^{i\phi_{\rm AB_{2}}(\nu)} \eta(\nu)  D_2(E)\ln{\left(\frac{E}{g_{\rm s_2 0}}\right)}\bigg]
\end{align}
In general, $\eta(\nu)$ is a complex number. Its argument quantifies the propagation time lag between the variations in each lamp post. This is the case for the two-blob model of \cite{Chainakun17}, and will be explored in \texttt{reltrans} in future work. Throughout this paper, we will instead assume that the two lampposts vary incoherently. In this case, the phase lag between the two lampposts has no effect on the predicted cross-spectrum. For the purposes of this paper, we therefore choose to set $\eta(\nu)$ to a real number.
\begin{figure*}
    \includegraphics[width=\textwidth , trim={0.65cm 0.0cm 0.6cm 0.0cm},clip]{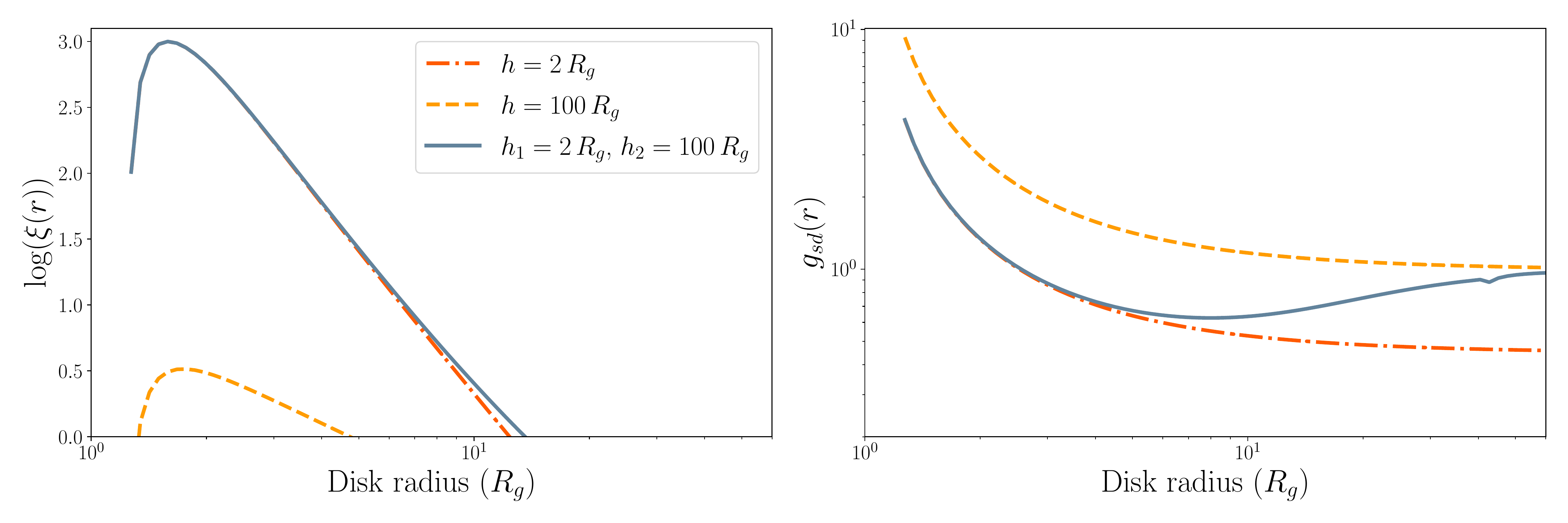}
    \caption{Left panel: radial scaling of the disk ionization calculated with $h_1=2\,\rm{R_g}$, $h_2=100\,\rm{R_g}$, and a peak of $\log(\xi)=3$. The red and yellow lines show the contribution from the lamp posts at each height, and the gray line shows the total profile for the double lamp post model. Right panel: radial scaling for the source-to-disk g-factor of single lamp post models, as well as the effective factor defined in Eq.~\ref{eq:g_eff} for the double source case. The color convention is identical to the left panel.}
    \label{fig:radial_scaling}
\end{figure*}
\subsection{Reflection}
\label{sec:reflect}

The inclusion of a second X-ray source introduces additional complexities when computing the reflection spectrum, which we detail in this section. First, each lamp post will contribute to the total ionization state of the disk, and secondly, the disk does not see a single exponentially cut-off illuminating spectrum, but rather a combination of two such spectra with different cutoffs. In general, these differences are subtle, and we expect them to only play a major role if the upper lamp post is significantly brighter than the lower one.

Generalizing eq. 14 from \cite{Ingram19} (and dropping the $\nu$ dependency as we only consider time-averaged quantities in this section), the specific irradiating flux crossing a disk patch at coordinates $r,\phi$ (calculated in Boyer-Lindquist coordinates) is:
\begin{align}
    F_{irr}(E_d,r,\phi)= C_1 \epsilon_1(r)E_d^{1-\Gamma}{\rm e}^{-E_d/(\rm g_{s_1 d} E_{\rm cut})} + \nonumber \\
    C_2 \epsilon_2(r)E_d^{1-\Gamma}{\rm e}^{-E_d/(g_{\rm s_2 d} E_{\rm cut})};
    \label{eq:flux_irr_spec}
\end{align}
$E_{\rm d}$ is the photon energy calculated in the frame of each disk patch and  $g_{\rm s_j d}$ is the energy shift experienced by photons traveling from the $j$-th lamp post to the disk. The emissivity profile for the $j-$th lamp post is
\begin{equation}
    \epsilon_j(r) = \frac{g_{s_j d}^{\Gamma}}{2} \frac{|d \cos\delta/dr|_j}{(dA_{\rm ring}/dr)};
    \label{eq:emis}
\end{equation}
$\delta$ is the angle between a photon emitted by one of the lamp posts and the spin axis of the black hole, measured in the rest frame of each X-ray source, $dr$ is the width of each patch of the disk located at radius $r$, and $dA_{\rm ring}$ is the area of each disk annular patch measured in its rest frame.

We can calculate the bolometric flux by integrating Eq.~\ref{eq:flux_irr_spec} from $E_d=0$ to $E_d=\infty$. Changing variables to $E$ via the relation $E_d = (g_{\rm s_j d}/g_{\rm s_j o})E$, and substituting $C_{2}=\eta_{0} C_{1}$, gives:
\begin{align}
    F_x(r)  = & C_1 \epsilon_1(r) \left(\frac{g_{\rm s_1 d}}{g_{\rm s_1 o}}\right)^{2-\Gamma} \Theta_1 + \nonumber \\ 
    & \eta_0 C_1 \epsilon_2(r) \left(\frac{g_{\rm s_2 d}}{g_{\rm s_2o}}\right)^{2-\Gamma} \Theta_2,
    \label{eq:flux_irr}
\end{align}
where
\begin{equation}
    \Theta_j \equiv \int_0^\infty E^{1-\Gamma} {\rm e}^{-E/ ( g_{\rm s_j o} E_{\rm cut} ) }dE.
\end{equation}
The ionization parameter $\xi(r) = 4\pi F_x(r) / n_e(r)$ (assuming that the electron and hydrogen number densities are equal) can then be calculated from Eq.~\ref{eq:flux_irr} by assuming a radial density profile $n_e(r)$. As in previous versions of \texttt{reltrans}, it is possible to either take $n_{e}(r)$ to be constant, or let it follow the Shakura-Sunyaev density profile \citep{Shakura73}, in which case $n_{e}(r) \propto r^{3/2}[1-(r_{\rm in}/r)^{1/2}]^{-2}$. We assume the Shakura-Sunyaev profile throughout this paper. Finally, to normalize the radial ionization profile without knowing the distance to the source a-priori, we can define:
\begin{align}
    \xi_*(r) \equiv \frac{1}{n_e(r)}\Biggl\{ \epsilon_1(r) \left(\frac{g_{\rm s_1 d}}{g_{\rm s_1 o}}\right)^{2-\Gamma} \Theta_1 + \nonumber \\ 
      \eta_0 \epsilon_2(r) \left(\frac{g_{\rm s_2 d}}{g_{\rm s_2 o}}\right)^{2-\Gamma} \Theta_2 \Biggl\},
\end{align}
such that the ionization parameter becomes
\begin{equation}
    \xi(r) = \xi_{\rm max} \frac{\xi_*(r)}{[\xi_*(r)]_{max}},
\end{equation}
where $\xi_{\rm max}$ is the highest value of the ionization parameter throughout the disk. The left panel of fig.\ref{fig:radial_scaling} shows an example radial ionization profile, with $h_1=2\,\rm{R_g}$, $h_{2}=100\,\rm{R_g}$, $\log(\xi_{max})=3\,\rm{\log(erg\,s^{-1}\,cm)}$, and $\eta_0=1$, representing a vertically extended corona, in which light bending effects cause a large fraction of the continuum signal to originate from the top of the X-ray source. For this set of parameters, only the lowest lamp post contributes to the disk ionization at small ($R\leq 10\,\rm{R_g}$) disk radii. The ionization profile deviates somewhat from the single lamp post case around $10\,\rm{R_g}$, where the second source starts to contribute. This behavior is purely a consequence of the lower X-ray source being much closer to the inner disk, resulting in a much higher ionizing flux. Instead, in the outer regions, the higher source may appear to be brighter in the disk frame, but its flux is low enough that the disk will remain un-ionized.

As discussed at the start of this section, we assume that the lamp posts have identical cutoff energies in their rest frames, and as a result the spectrum seen by each disk patch is not just a cutoff power-law, but a superposition of two such spectra. This change in continuum spectral shape should introduce subtle changes to the reflection spectrum; however, available reflection models (e.g. \texttt{xillver} or \texttt{reflionx}) assume that the illuminating spectrum is a single power-law or Comptonisation component, rather than a superposition of two such components (with potentially different photon indexes). We address this issue by defining an effective source-to-disk g-factor, as the luminosity-weighted average of the g-factor from each source:
\begin{equation}
    g_{\rm sd, eff}(r) = \frac{F_{x,1}(r) g_{\rm s_1 d}(r) + F_{x,2}(r) g_{\rm s_2 d}(r)}{F_{x,1}(r)+F_{x,2}(r)},
    \label{eq:g_eff}
\end{equation}
where $F_{x,j}(r)$ is the flux from the j-th source seen by each disk radius $r$. We then set the energy of the cutoff in \texttt{xillver} as $E_{\rm cut, disk} = g_{\rm sd, eff} E_{\rm cut}$ when computing the reflection spectrum in the rest-frame of the disc. A comparison of $g_{\rm sd, eff}$  with two single lamp post cases is shown in the right panel of fig.\ref{fig:radial_scaling}. Intuitively, $g_{\rm sd, eff}$ is nearly identical to the g-factor from the lower/upper lamp posts in the inner/outer disk, respectively, where each of the two sources dominates, and takes an intermediate value between these two extremes.

The irradiating flux is all reprocessed in the reflection spectrum, and therefore, in analogy with eq. 19 in \cite{Ingram19}, the reflected flux from each patch of the disk is:
\begin{align}
    dR(E,r,\phi) = C_1[\epsilon_1(r) + \eta_0\epsilon_2(r)]\times \nonumber \\
    g_{\rm do}^{3}(r,\phi)\mathcal{R}(E/g_{\rm do}|\mu, g_{\rm sd, eff} E_{\rm cut}, \xi, n_{\rm e}) d\alpha d\beta. 
    \label{eq:reflection_spectrum}
\end{align} 
$\mathcal{R}$ is the reflection spectrum calculated in the rest-frame of the disk, $g_{\rm d o}$ is the disk-to-observer energy shift of the photons, and $\alpha$ and $\beta$ are the impact parameters of the photons on the observer plane. From Eq.~\ref{eq:reflection_spectrum}, one can clearly see that $\eta_0$ affects not just the relative importance of the two continua, but also the relative importance of the reflection driven by each. As a result, in the limit of $\eta_0\to 0$ we recover the single lamp post case with $h=h_1$, while for $\eta_0\to \infty$ the spectrum corresponds to that of a single source with $h=h_2$. Finally, note that Eq.~\ref{eq:reflection_spectrum} is not equal to the sum of two reflection spectra in the lamp post geometry with heights $h_1$ and $h_2$ because both the radial dependency of the ionization and cutoff energy seen by each disk patch is different from any single lamp post model. We discuss this difference further in Sec.\ref{sec:behavior}.

\begin{figure}
    \centering
    \includegraphics[width=\columnwidth, trim={0.65cm 0.0cm 0.4cm 1.5cm},clip]{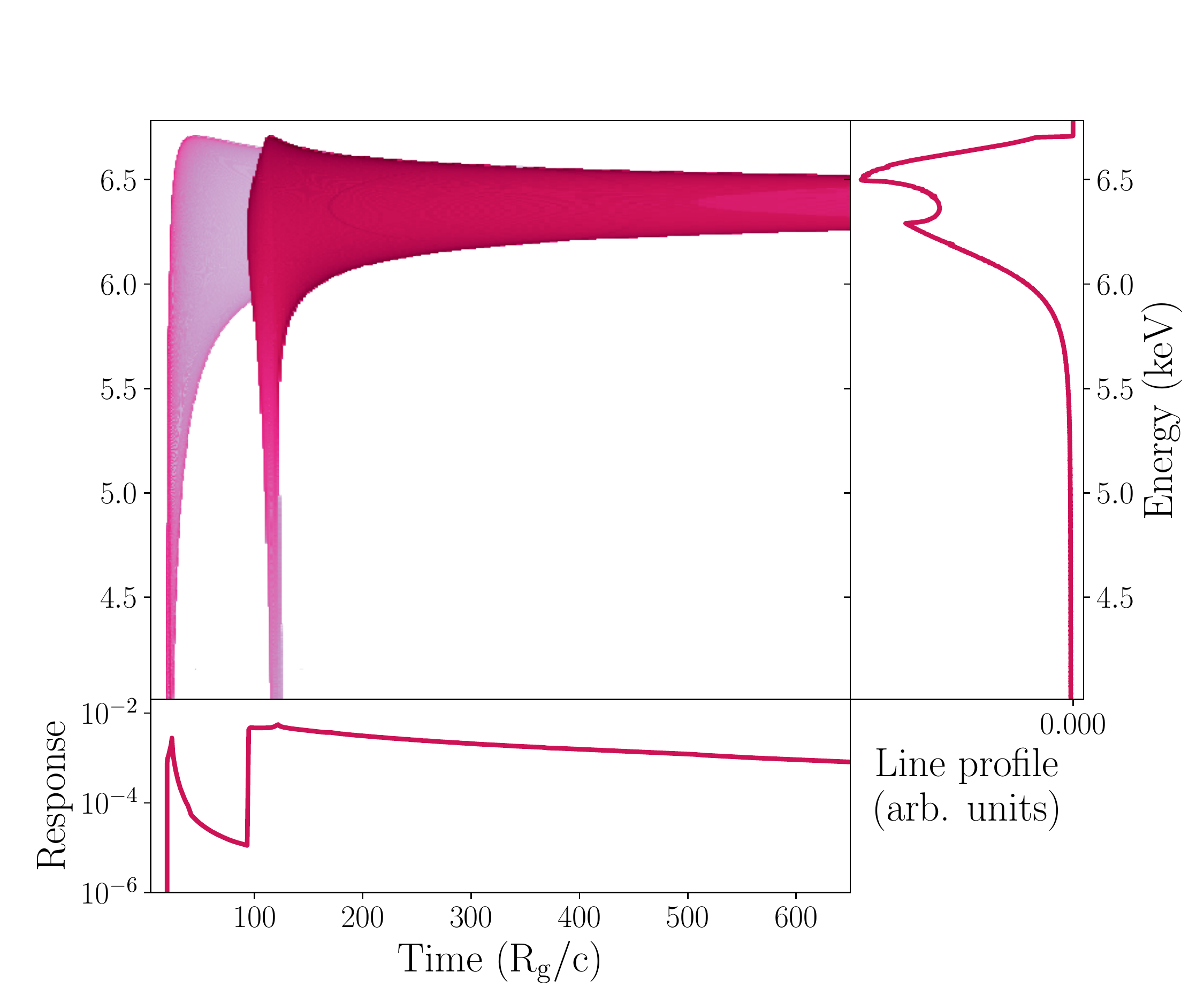}
    \caption{Impulse response function kernel for the double lamp post model to a mono-energetic flash of $6.4$ keV photons. In the center panel, lighter/darker colors indicate a smaller/larger reflection signal respectively, at a given time and energy. The bottom panel shows the disk response over time, integrated over all energies. The right panel shows the energy-dependent disk response integrated over all times, which corresponds to the iron line profile.}
    \label{fig:impulse}
\end{figure}

\subsection{Reverberation}
\label{sec:reverb}

In the single lamp post case, the general form of the total time-dependent reflection signal is $R(E,t) = C(t)\otimes w(E,t)$, where $\otimes$ stands for a convolution and:
\begin{align}
w(E,t) = \int_{\alpha,\beta}\epsilon(r)g^{3}_{\rm do}(r,\phi)\delta(t-\tau(r,\phi)) \nonumber \\
        \times\mathcal{R}(E/g_{\rm do}|\d  E_{\rm cut}, \xi) d\alpha d\beta 
    \label{eq:imuplse_single_time}
\end{align}
is the disk response function. In the time dependent case,  $R(E,t)$ is evaluated at a time $t-\tau(r,\phi)$, with $ \tau(r,\phi) = \tau_{\rm s d}(r) + \tau_{\rm do}(r,\phi) - \tau_{\rm s o}$; $\tau_{\rm do}$ is the light travel time from the disk to the observer and $\tau_{\rm s o}$ is the light travel time from the point source to the observer. In order to save computational time, it is convenient to express Eq.~\ref{eq:imuplse_single_time} in Fourier space, such that the convolution becomes a product: $R(E,\nu) = C(\nu) W(E,\nu)$, where:
\begin{align}
    W(E,\nu) = \int_{\alpha,\beta}\epsilon(r)g^{3}_{\rm do}(r,\phi) e^{i2\pi t(r,\phi)\nu}  \nonumber \\
    \times\mathcal{R}(E/g_{\rm do}|\mu, E_{\rm cut}, \xi)  d\alpha d\beta.
    \label{eq:imuplse_single_freq}
\end{align}
is the disk transfer function. In order to further optimize the code, Eq.~\ref{eq:imuplse_single_freq} is evaluated in \texttt{reltrans} by convolving the rest-frame reflection spectrum with a kernel, and changing variables from $E$ to $\log E$. In this way, $W(\log(E),\nu) = \mathcal{R}(\log(E))\otimes W_{\delta}(\log(E),\nu)$, or more explicitly: 
\begin{align}
    W(\log(E),\nu) = \int_0^{\infty }\mathcal{R} (\log(E^{\prime}) \times \nonumber \\
    W_{\delta } (\log(E/E^{\prime}),\nu) d\log(E^{\prime})
\end{align}
and the kernel $W_{\delta}(\log(E/E^{\prime},\nu)$ is given by:
\begin{figure}
    \includegraphics[width=\columnwidth, trim={0.65cm 0.0cm 0.6cm 0.0cm},clip]{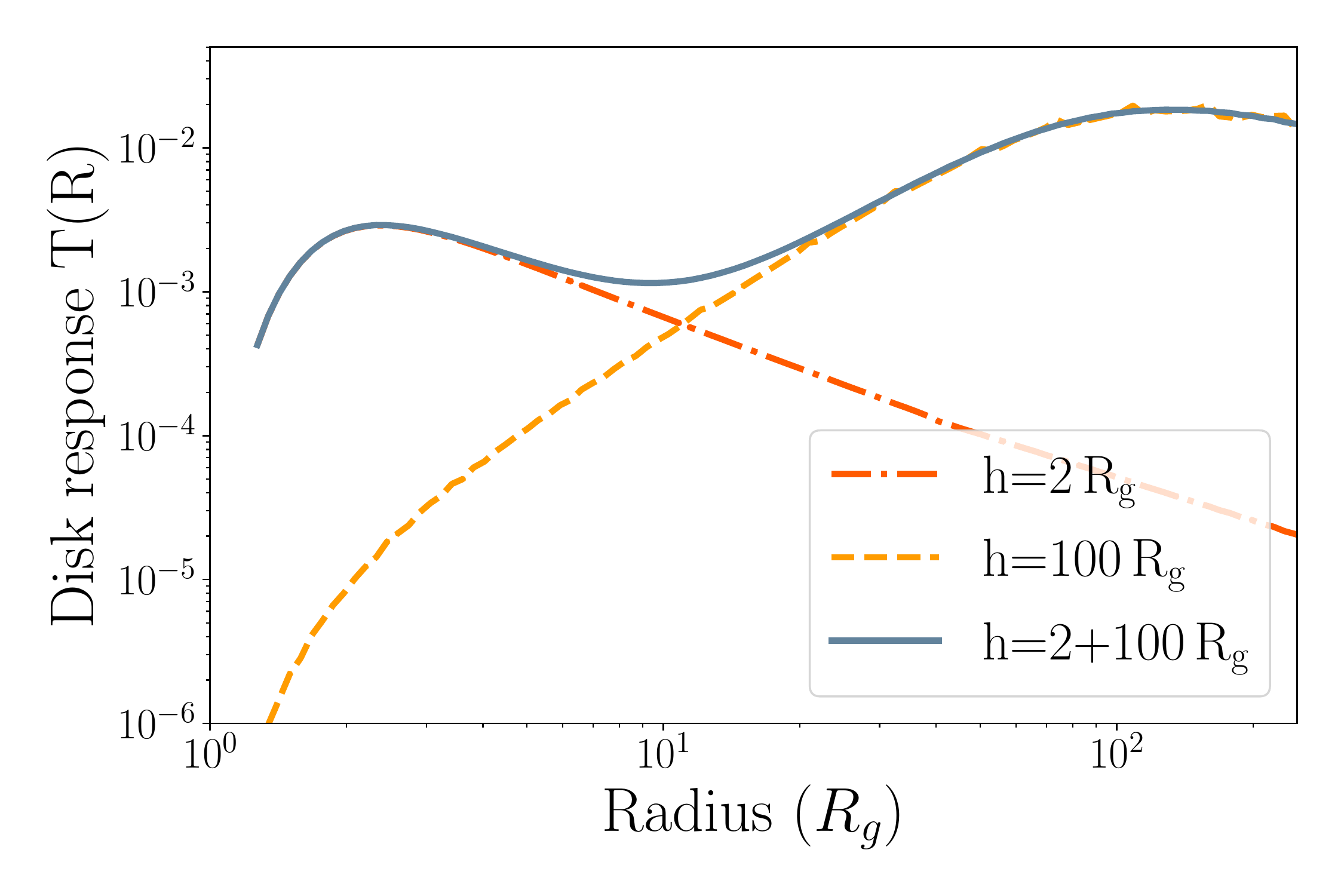}
    \caption{Radial dependency of the disk response function kernel, integrated over time and energy. As in fig.\ref{fig:radial_scaling}, the red and yellow lines correspond to the signals driven by the low/high lamp posts, which originate respectively at small/large radii, and the gray line shows the total disk response for the double lamp post model.}
    \label{fig:radial_impulse}
\end{figure}
\begin{align}
    W_{\delta}(\log(E/E^{\prime},\nu) = \int_{\alpha,\beta}\epsilon(r)g^{3}_{\rm do}(r,\phi)e^{i2\pi \tau(r,\phi)\nu}\nonumber \\
    \times \delta\left[\log(E) - \log\left( g_{\rm do}\right)  \right] d\alpha d\beta 
    \label{eq:transfer_func}
\end{align}
Eq.~\ref{eq:transfer_func} encodes the response of the disk to a mono-energetic flash of light, in Fourier space. 

In the double lamp post case, the kernel from each source encodes the disk response to a mono-energetic flash, emitted at each source height. Its expression for the $j$-th source becomes:
\begin{align}
    W_{\delta}(\log(E/E^{\prime}),\nu) = & \int_{\alpha,\beta} \epsilon_j(r)g^{3}_{\rm do}(r,\phi)e^{i2\pi \tau_j(r,\phi)\nu}  \nonumber \\
    &\times \delta\left[\log(E) - \log (g_{\rm do}) \right] d\alpha d\beta 
    \label{eq:transfer_double};
\end{align}
in this case, we simply compute the kernel for each lamp-post. We then convolve each kernel with the rest-frame reflection spectrum, obtaining the transfer functions encoding the reverberation signal from each lamp post.

Plotting both these kernels together is useful to highlight the changes introduced by the addition of a second lamp post. One such example is shown in fig.\ref{fig:impulse}; we consider the same case as previous figures, with $h_1=2\,R_{\rm g}$, $h_2=100\,R_{\rm g}$, and $\eta(\nu) = 1$; other parameters are dimensionless spin $a^{*}=0.998$, innermost disk radius $R_{\rm in}=R_{\rm ISCO}$ (where $R_{\rm ISCO}$ is the radius of the innermost circular orbit) and inclination $\rm{Incl}=30^{\circ}$. Intuitively, the response to the upper lamp post is greatly delayed with respect to the lower one. The total line profile shows only moderate relativistic broadening, indicating that the contribution from the top of the corona dominates, and includes three main features: two narrow peaks near $6.2-6.5\,\rm{keV}$, caused by the Doppler effects in the outer disk, and a weak broad component extending down to $\approx 4\,\rm{keV}$, driven by the lower lamp post. The latter signal dominates at early times, when the reflected photons originate near the black hole, as one would expect. 

Finally, in this case (which corresponds to taking $\eta(\nu)=1$) it appears that the total signal is dominated by the top of the corona. This behavior can be understood by plotting the radial dependency of the disk response, which quantifies how much each disk radius contributes to the total reflected spectrum. This function is plotted in fig.\ref{fig:radial_impulse}. One can immediately see that the signal is dominated by the source with the larger height, and produced in the outer regions of the disk. Compared to the smaller radii illuminated by the bottom lamp post, these regions have much larger surface area and experience less gravitational redshift, resulting in a larger reflected signal. This behavior also explains why the narrow component of the iron line is so prominent over the broad one, as shown in fig.\ref{fig:impulse}. This example also shows how $\eta(\nu)$ behaves: it effectively acts as a way to tweak the relative normalization of the two kernels/transfer functions.

Beyond pure light-travel time delays, \texttt{reltrans} accounts for further lags driven by the pivoting of the reflection spectrum, as well as the changes in the ionization state of the disk, driven by the pivoting of the continuum. Identically to previous model flavors \citep[e.g.][and references therein]{Mastroserio21}, these variations are accounted for with a first order Taylor expansion of the reflection spectrum. As a result the total reflection signal $R(E,\nu)$ can be expressed as a sum of individual transfer functions, each accounting for a specific mechanism driving its own lags. In total, this version of the model computes eight transfer functions in the reflection term, four for each lamp post. Of these, two encode light-travel time delays ($W_{0,1}$ and $W_{0,2}$) from each source, two the changes in emissivity profile ($W_{1,1}$ and $W_{1,2}$) and reflection spectrum ($W_{2,1}$ and $W_{2,2}$) due to the pivoting of the photon index, and two the variations in the ionization state of the disk as the continuum luminosity varies ($W_{3,1}$ and $W_{3,2}$). The full derivation of each term in the time-dependent reflection spectrum is presented in Appendix A. The transfer functions accounting for light travel time delays are given by: 
\begin{align}
W_{0,1}(E,\nu) = \int_{\alpha,\beta} g_{\rm do}^3 \epsilon_1(r) {\rm e}^{i 2\pi \tau_{\rm R_1} \nu} \mathcal{R}(E_d)~d\alpha d\beta, \nonumber \\
W_{0,2}(E,\nu) = \int_{\alpha,\beta} g_{\rm do}^3 \epsilon_2(r) {\rm e}^{i 2\pi \tau_{\rm R_2} \nu} \mathcal{R}(E_d)~d\alpha d\beta, \label{eq:w0} 
\end{align}
Those accounting for the pivoting reflection spectrum are:
\begin{align}
W_{1,1}(E,\nu) = \int_{\alpha,\beta} g_{\rm do}^3 \epsilon_1(r) ~\ln g_{\rm s_1 d} ~{\rm e}^{i 2\pi \tau_{\rm R_1} \nu} ~\mathcal{R}(E_d)~d\alpha d\beta, \nonumber \\
W_{1,2}(E,\nu) = \int_{\alpha,\beta} g_{\rm do}^3 \epsilon_2(r) ~\ln g_{\rm s_2 d} ~{\rm e}^{i 2\pi \tau_{\rm R_2} \nu} ~\mathcal{R}(E_d)~d\alpha d\beta, \nonumber 
\end{align}
\begin{align}
W_{2,1}(E,\nu) = \int_{\alpha,\beta} g_{\rm do}^3 \frac{C_{1,0} \epsilon_1(r) + C_{2,0} \epsilon_2(r)}{C_{1,0}+C_{2,0}} \times \nonumber \\
{\rm e}^{i 2\pi \tau_{\rm R_1} \nu}\frac{\partial \mathcal{R}(E_d)}{\partial \Gamma}~d\alpha d\beta,  \nonumber \\
W_{2,2}(E,\nu) = \int_{\alpha,\beta} g_{\rm do}^3 \frac{C_{1,0} \epsilon_1(r) + C_{2,0} \epsilon_2(r)}{C_{1,0}+C_{2,0}} \times \nonumber \\
{\rm e}^{i 2\pi \tau_{\rm R_2} \nu} \frac{\partial \mathcal{R}(E_d)}{\partial \Gamma}~d\alpha d\beta.
\end{align}
and those capturing the changes in disk ionization are
\begin{align}
W_{3,1}(E,\nu) = \frac{1}{\ln 10} \int_{\alpha,\beta} g_{\rm do}^3 \kappa(r) \Theta_1(r) \times \nonumber \\
{\rm e}^{i 2\pi \tau_{\rm R1} \nu} \frac{\partial \mathcal{R}(E_d)}{\partial\log\xi}~d\alpha d\beta, \nonumber\\
W_{3,2}(E,\nu) = \frac{1}{\ln 10} \int_{\alpha,\beta} g_{\rm do}^3 \kappa(r) \Theta_2(r)  \times \nonumber \\
{\rm e}^{i 2\pi \tau_{\rm R2} \nu} \frac{\partial \mathcal{R}(E_d)}{\partial\log\xi}~d\alpha d\beta,
\label{eq:w3}
\end{align}
where the terms $\kappa(r)$ and $\Theta_i(r)$ are defined in Appendix A. Furthermore, the rest-frame reflection spectrum in equations \ref{eq:w0}--\ref{eq:w3} depends on the ionization parameter $\xi$, the source-to-disk g factor $g_{\rm sd, eff}$, the disk density $n_{\rm e}$, and the trajectory $\mu_{\rm e}$ of the photons that emerge from the disk and reach the observer, all of which vary along the disk radius. This is accounted for by dividing the disk into $K$ radial bins for the ionization, g-factor and density, and $J$ bins for the photon trajectory $\mu_{\rm e}$; the $i-$th transfer function is calculated by summing each disk patch contribution over all bins:
\begin{equation}
    W_{i}(E,\nu) = \sum_{j=1}^{J} \sum_{k=1}^{K} \Delta W_{i}(E,\nu | \mu_{\rm e}(j), \xi(k), g_{\rm sd, eff}(k), n_{\rm e}(k))
\end{equation}
Finally, we can finally add the direct and reflected components for each source to get:
\begin{align}
S_1(E,\nu) = C_1(\nu) \bigg\{ D_1(E) + \frac{W_{0,1}(E,\nu)}{\mathcal{B}}  \nonumber \\ 
+ \frac{W_{3,1}(E,\nu)}{\mathcal{B}} + \gamma_1(\nu) {\rm e}^{i \phi_{AB,1}(\nu)} \bigg[ \frac{W_{1,1}(E,\nu)}{\mathcal{B}} \nonumber \\
+ \frac{W_{2,1}(E,\nu)}{\mathcal{B}}  - D_1(E) \ln(E/g_{\rm s_1 o}) \bigg] \bigg\}, 
\label{eq:fullmodel1}
\end{align}
\begin{align}
S_2(E,\nu) =  C_2(\nu)\bigg\{ {\rm e}^{i 2\pi \tau_{d} \nu} D_2(E) + \frac{W_{0,2}(E,\nu)}{\mathcal{B}}  \nonumber \\
+ \frac{W_{3,2}(E,\nu)}{\mathcal{B}} + \gamma_2(\nu) {\rm e}^{i \phi_{AB,2}(\nu)} \bigg[  \frac{W_{1,2}(E,\nu)}{\mathcal{B}} \nonumber \\
+ \frac{W_{2,2}(E,\nu)}{\mathcal{B}} - {\rm e}^{i 2\pi \tau_{d} \nu} D_2(E) \ln(E/g_{\rm s_2 o}) \bigg] \bigg\}.
\label{eq:fullmodel2}
\end{align}
where the factor $1/\mathcal{B}$, called the boost parameter, acts as an adjustment to the reflection fraction from that calculated self-consistently in the model assuming isotropic and stationary lamp post sources. The model lag-energy spectrum is computed similarly to the procedure described in \cite{Mastroserio20}, starting from eq.\ref{eq:fullmodel1} and \ref{eq:fullmodel2}. Because the two sources are assumed to be fully incoherent, we can calculate the reference band count rate from each source separately, compute the cross spectrum separately, and then sum the two contributions to obtain the total model. We compute the Fourier transform of the reference band $R_{r,j}(\nu)$ for the $j$-th lamp post self consistently from the model:
\begin{equation}
    R_{j,r}(\nu) = \sum_{I=I_1}^{I=I_2} S_j(I,\nu),
\end{equation}
where $I_1$ and $I_2$ are the lowest and highest energy channels of the reference band, and $I$ those in-between. With this notation, the cross-spectrum for each source is is:
\begin{align}
   G_j(E,\nu) & = S_j(E,\nu) R^{*}_{j,r}(\nu) \\ \nonumber
   & = |C_{j}(\nu)|^{2} S_{j,\rm raw} (E,\nu) R^{*}_{j,\rm raw}(\nu) \\ \nonumber
   & = |C_{j}(\nu)|^{2} G_{j,\rm raw}(E,\nu),
   \label{eq:cross_spectrum}
\end{align}
where  $G_{j,\rm raw}(E,\nu) = S_{j,\rm raw}(E,\nu)R^{*}_{j,\rm raw}(\nu)$, $S_{j,\rm raw}(E,\nu) = S_j(E,\nu)/C_j(\nu)$ and $R^{*}_{j,\rm raw}(\nu) =  R_{j,r}(\nu)/C_{j}(\nu)$. The total model then is simply the sum of the two contributions:
\begin{equation}
    G_{\rm sum}(E,\nu_{\rm c}) = |C_1(\nu)|^2 \big[ G_{1,\rm raw}(E,\nu) + \eta^2(\nu)G_{2,\rm raw}(E,\nu) \big]
   \label{eq:G_avg} 
\end{equation}
Finally, following \cite{Mastroserio21}, an additional phase term $e^{i\phi_{\rm A}(\nu)}$ is included to account for instrument calibration:
\begin{equation}
    G_{\rm raw}(E,\nu_{\rm c}) = e^{i\phi_{\rm A}(\nu)}G_{\rm sum}(E,\nu_{\rm c})
\end{equation}
From the cross spectrum, the model lag-energy spectrum (and energy-dependent cross spectral amplitude) can be computed by averaging over a frequency range from $\nu_{\rm lo}$ to $\nu_{\rm hi}$, centered at $\nu_{\rm c} = (\nu_{\rm hi}+\nu_{\rm lo})/2$ (and assuming that the model parameters are constant over this frequency range):
\begin{align}
    \langle G(E,\nu_{\rm c}) \rangle = \int_{\nu_{\rm lo}}^{\nu_{\rm hi}} \frac{e^{i\phi_{\rm A}(\nu)}|C_{1}(\nu)|^{2}}{\nu_{\rm hi}-\nu_{\rm lo}}\times \nonumber \\ 
    \big[ G_{1,\rm raw}(E,\nu) + \eta^2(\nu)G_{2,\rm raw}(E,\nu) \big] d\nu
\end{align}

In order to compute Eq.~\ref{eq:G_avg}, we need to assume a frequency dependence for $|C_{1}(\nu)|^{2}$. We note that in the single lamp post case, this quantity roughly corresponds to the PSD of the broadband noise \citep{Mastroserio20}; as a result, previous versions of the model assume it is a power-law with slope $k=-2$. The reason for this choice is that at low Fourier frequencies ($\leq 5\,\rm{Hz}$, where the PSD of accreting black holes generally has slopes of either $k=0$ or $k=-1$), the model is insensitive to the choice of $k$. At high Fourier frequencies, instead, the broadband noise slope is roughly $k=-2$; as a result, this value is taken for every Fourier frequency for simplicity \citep{Mastroserio20,Mastroserio21}. However, in the double lamp post case, $|C_{1}(\nu)|^{2}$ only carries information about the rms of the lower lamp post. In order to avoid over-complicating the model assumption we use the same $k=-2$ power-law form as previous flavors of the model:
\begin{equation}
 |C_{\rm 1}(\nu)|^{2} = \frac{\alpha(\nu_{\rm c})}{1+\eta(\nu_{\rm c})}\left(\frac{\nu}{\nu_{\rm c}}\right)^{-2};  
\end{equation}
including the factor $1+\eta(\nu_{\rm c})$ is purely for convenience, as with this choice when $\eta(\nu_{\rm c})$ is set to 0 the model normalization is identical to the single lamp post case. As a result of these additional changes and assumptions, we note that $\alpha(\nu_c)$ and $ |C_{\rm 1}(\nu)|^{2}$ are not as easily interpreted as in older flavors of the model. The frequency-averaged cross spectrum for each source is:
\begin{align}
    \langle G(E,\nu_{\rm c}) \rangle = \frac{\alpha(\nu_{\rm c})}{1+\eta(\nu_{\rm c})} \int_{\nu_{\rm lo}}^{\nu_{\rm hi}} \frac{e^{i\phi_{\rm A}(\nu)}(\nu/\nu_{\rm c})^{-2}}{\nu_{\rm hi}-\nu_{\rm lo} } \times \nonumber \\
     \big[ G_{1,\rm raw}(E,\nu) + \eta^2(\nu)G_{2,\rm raw}(E,\nu) \big] d\nu
    \label{eq:final_cross},
\end{align}
and as standard, the time lag is computed from Eq.~\ref{eq:final_cross} as $\tau_{\rm lag}(E,\nu_{\rm c}) = \arg[G(E,\nu_{\rm c})]/2\pi\nu_{\rm c}$. Finally, we account for neutral absorption including the \texttt{tbabs} model from \texttt{Xspec}, and for the source redshift in the same way as in older model flavors. A summary of the model parameters is given in Appendix B, together with the model parameter values used in this and the following section. 

\section{Model behavior}
\label{sec:behavior}

In this section, we discuss explicitly the impact that the addition of a second X-ray source of varying luminosity has on both time-averaged and lag-energy spectra. Unless explicitly stated, we set the model parameters to the values quoted in Table \ref{tab:model_parameters}.

\begin{figure}
    \includegraphics[width=\columnwidth, trim={0.65cm 0.0cm 0.6cm 0.0cm},clip]{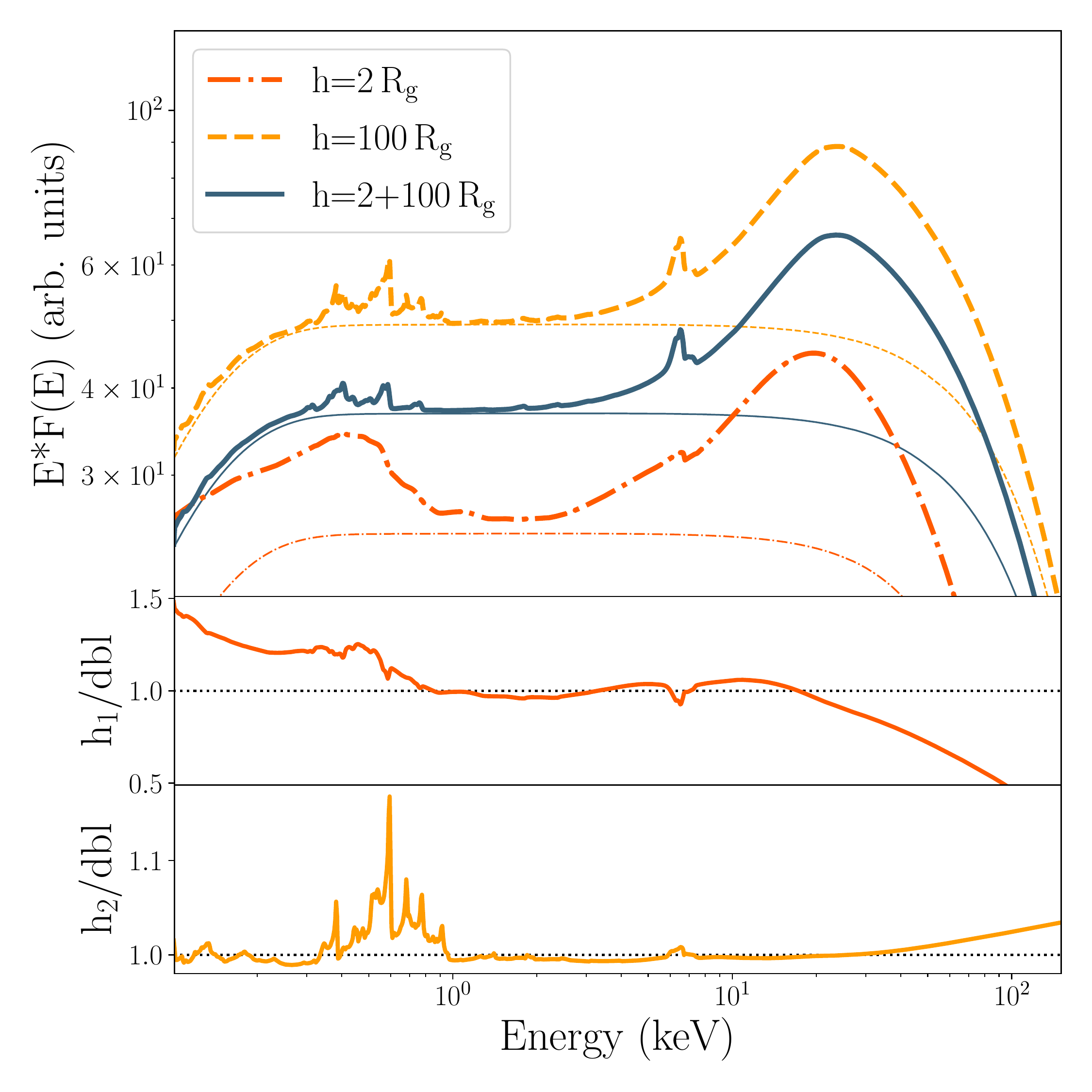}
    \caption{Top panel: comparison of time-averaged spectra, between single lamp post models with $h=2\,\rm{R_g}$, $h=100\,\rm{R_g}$ (red/yellow lines), and a double lamp post model with $h_1=2,\,h_2=100,\,\rm{R_g},\,\eta_{\rm 0}=1$ (blue line). The dashed lines indicated the continuum spectra alone. Middle and bottom panels: ratio between the spectra from each single lamp post case, and the double lamp post model. For the bottom two panels, the spectra have been re-normalized in order to have identical fluxes at 2 keV.}
    \label{fig:spectra_comparison}
\end{figure}
\begin{figure}
    \includegraphics[width=\columnwidth, trim={0.65cm 0.0cm 0.6cm 0.0cm},clip]{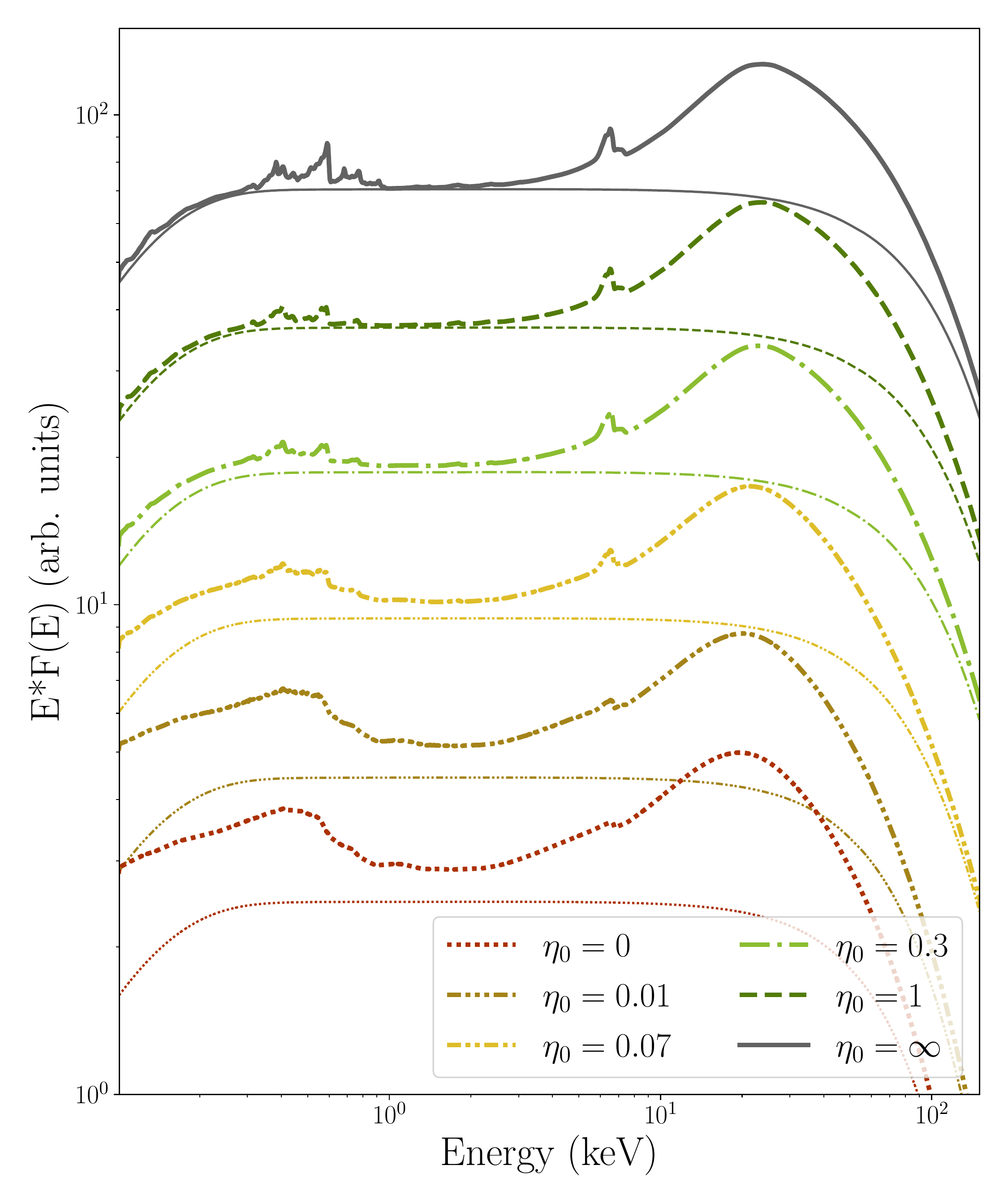}
    \caption{Comparison of time-averaged spectra with identical coronal parameters, as a function of $\eta_0$. The red line ($\eta_0=0$) is identical to a single lamp post model with $h=2\,\rm{R_g}$, and the gray line (($\eta_0=\infty$)) is identical to one with $h=100\,\rm{R_g}$. All the spectra have been shifted by an arbitrary multiplicative factor for clarity.}
    \label{fig:spectra_eta}
\end{figure}
Fig.\ref{fig:spectra_comparison} shows a comparison between the time-averaged spectra of the double lamp post model, computed with $\eta=1$, and both single lamp post cases. The double lamp post model generally looks similar to the large height case, rather than the low height one. This behavior is not surprising: for $\eta=1$ a large part of the signal originates in the top of the corona and is reflected in the outer disk, while for the low height source most of the radiation is reflected in the inner disk, where the total flux is suppressed by a combination of gravitational redshift and small effective area of the reflector. In particular, the shape of the iron line is very narrow and peaked, and almost identical to the large height single source case, as shown clearly in the bottom panels of the figure. A weak broad component can be seen in the ratio plots, but it is far less prominent than in the low height model. As a result, our model can in principle produce both (blurred) relativistic and (un-blurred) distant reflection components. The soft excess shows more complex behavior that is different from either single source model. On the one hand, the limited amount of relativistic smearing of the lines causes these features to overlap more and be less pronounced than the large height case; on the other, this effect is insufficient to completely blur the lines together and produce a smooth continuum, as in the highly relativistic model. 

Finally, the difference in the shape of the Compton hump and continuum cutoff is driven by two factors. Firstly, in each case the reflection in the rest frame is computed with a different cutoff through Eq.~\ref{eq:g_eff}. Secondly, we define the temperature in the frame of each lamp post, rather than in that of the observer, and then add the contribution of both continua. This choice naturally affects the shape of the high energy cutoff because of the different energy shifts from each lamp post to the observer. 

Fig.\ref{fig:spectra_eta} shows how the time-averaged spectra vary as a function of $\eta_{0}$. Small values of $\eta_{0}$ mean that the total (continuum+reflection) flux is dominated by the low height corona, and vice versa. Notably, the spectra differ significantly from the $h=2\,\rm{R_g}$ case even for fairly small values of $\eta_0$, as shown by the yellow line $\eta_{0}=0.07$: even if only a small fraction of the time-averaged signal originates in the larger lamp post, the iron line shows a noticeably narrow component, and individual features in the soft excess start becoming noticeable. Extremely small values of $\eta_0\approx 0.01$, on the other hand, are almost identical to the low height single source case. As a result, the parameter $\eta_0$ essentially controls the relative importance of the relativistic and distant reflection. 

\begin{figure}
    \includegraphics[width=\columnwidth, trim={0.65cm 0.0cm 0.6cm 0.0cm},clip]{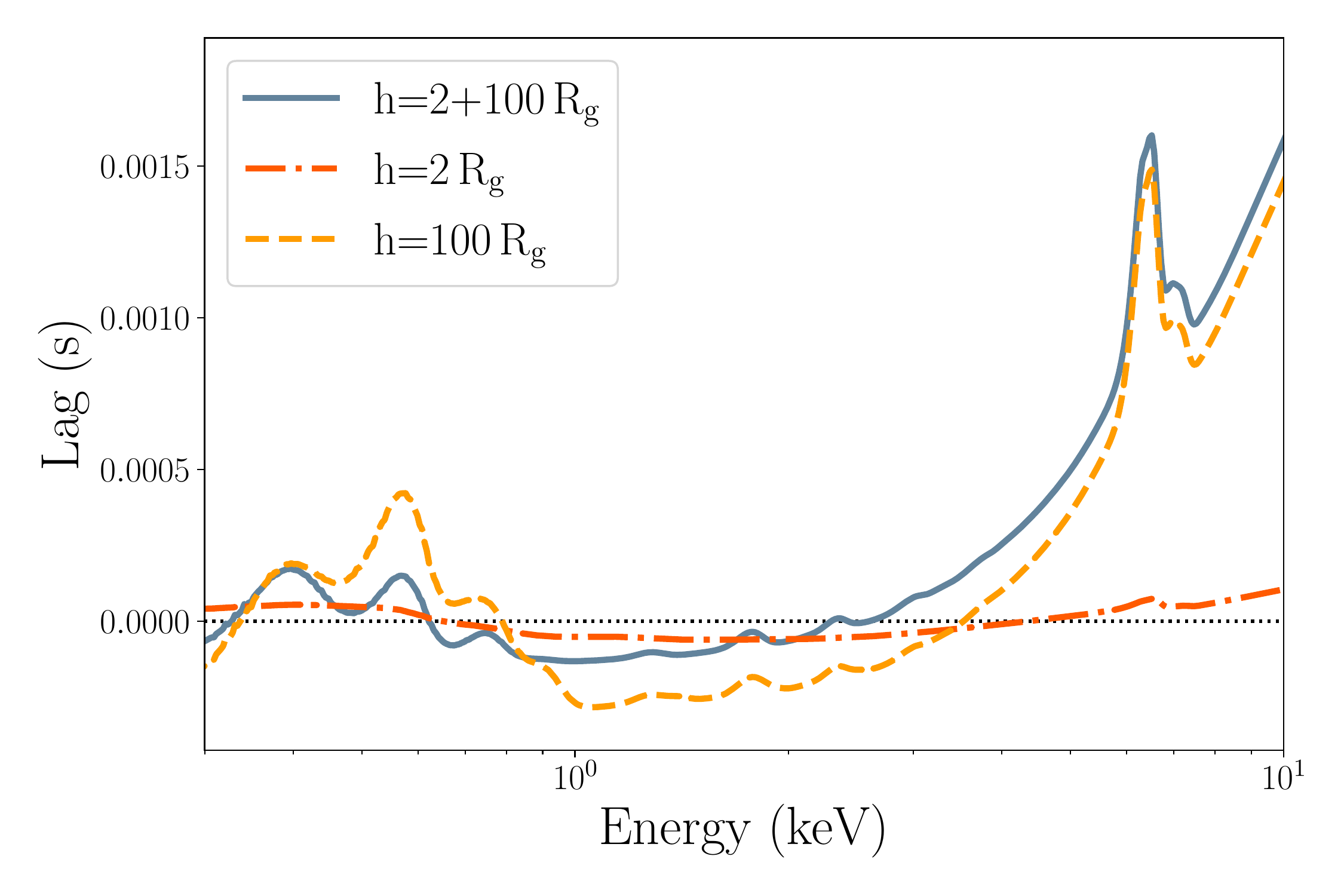}
    \caption{Comparison of lag-energy spectra (in the 1-10 Hz frequency range), including only the reverberation signal, for both single and double lamp post models. The color convention is the same as in previous plots.}
    \label{fig:lags_comparison}
\end{figure}

Fig.\ref{fig:lags_comparison} shows a comparison of the lag-energy spectra for the same models, focusing on the $0.3-10\,\rm{keV}$ band (which we take as the reference band), again with $\eta=1$ and neglecting the pivoting power-law. The model behavior is generally similar to the time-averaged spectra: the shape of the spectrum is most similar to a single lamp post model with large height. The amplitude of the lag is similar, but at low ($\leq 1\,\rm{keV}$) energies it is much weaker in the double lamp post case. This behavior is to be expected: the time-averaged spectra show a weaker soft excess (bottom panel of Fig.\ref{fig:spectra_comparison}), meaning that the low-energy reflection signal (driving the reverberation) is diluted with respect to the continuum. The dilution in these bands is thus more prevalent than for either single lamp post model, which results in a smaller lag below 1 keV.

\begin{figure}
    \includegraphics[width=\columnwidth, trim={0.65cm 0.0cm 0.6cm 0.4cm},clip]{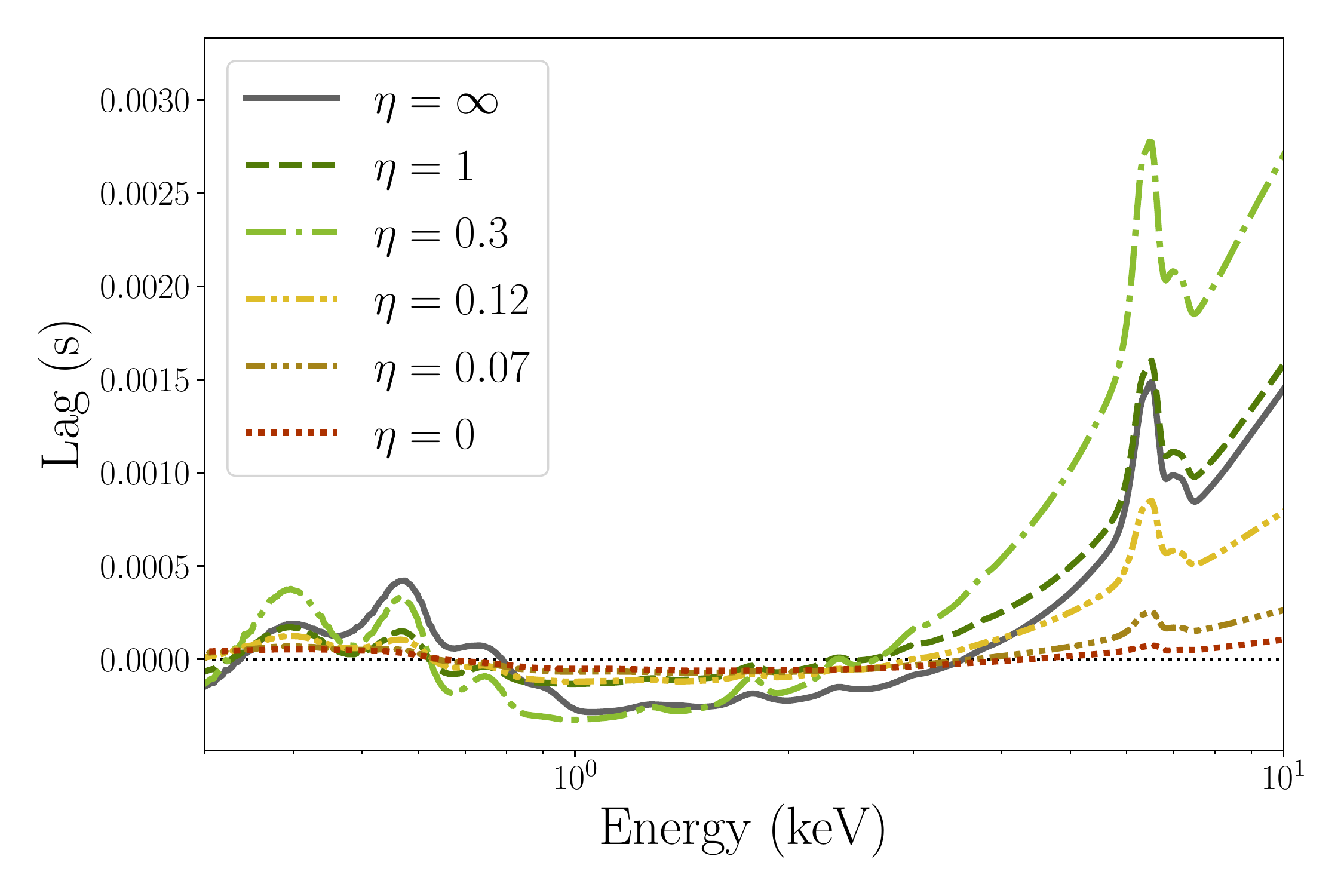}
    \caption{Comparison of lag-energy spectra (in the 1-10 Hz frequency range), showing only the reverberation signal, in a double lamp post model as a function of $\eta$. As in fig.\ref{fig:spectra_eta}, the red and gray line are analogous to single lamp post models with small/large heights, respectively.}
    \label{fig:lag_eta}
\end{figure}

Fig.\ref{fig:lag_eta} illustrates the effect of $\eta$ on the lag-energy spectra. While the overall behavior is similar to the time-averaged spectra, there are two crucial differences. First, even for small values of $\eta\approx 0.07$ the spectrum is noticeably different from the low height case, with the lag amplitude around the iron line increasing by a factor of $\approx 3$ compared to the low height single source case. Second, the lag amplitude above $1\,\rm{keV}$ increases noticeably for moderate values of $\eta=0.3$ (in which case the reverberation lag is larger than the single lamp post case), before decreasing for larger values. Intuitively, one would expect that larger values of $\eta$, corresponding to a larger fraction of the signal originating in the top corona, would lead to larger reverberation lags. However, this behavior is once again offset by dilution: larger values of $\eta$ also result in lower reflection fractions, as shown in fig.\ref{fig:spectra_eta}, and the total lag amplitude is a combination of light travel times and dilution. How these two phenomena balance each other is also evident at energies below $1\,\rm{keV}$: for all models, the lag is longer at energies with the most noticeable features in the time-averaged spectra. For example, the prominence of the line present in the time-averaged spectra at $0.5\,\rm{keV}$ also results in a similar feature in the lag spectra, with the single-lamp post case having a more prominent line (meaning dilution is lessened) and thus a longer lag. 

\begin{figure}
    \includegraphics[width=\columnwidth, trim={0.65cm 0.0cm 0.6cm 0.0cm},clip]{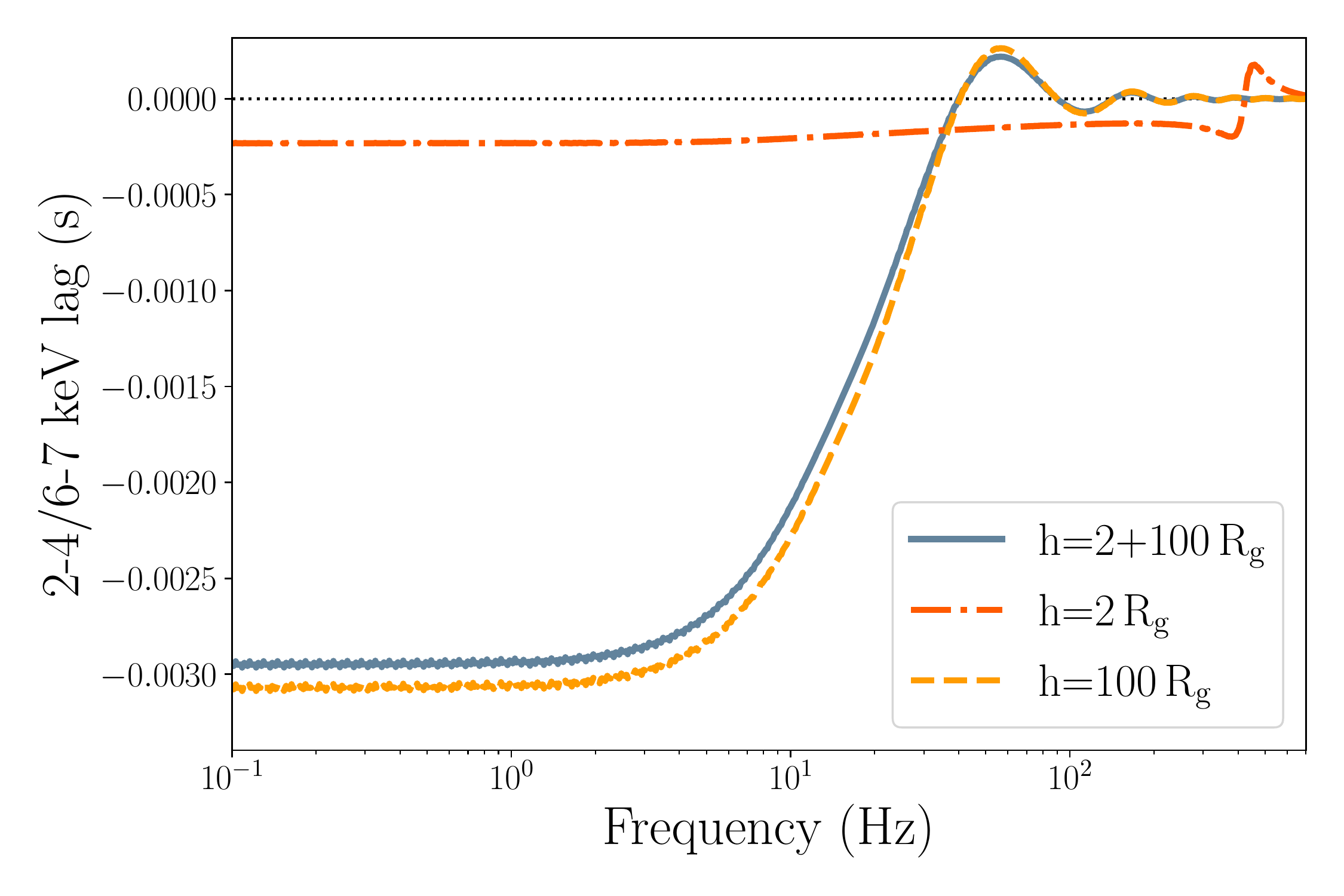}
    \caption{Comparison of lag-frequency spectra (without lags driven by the pivoting continuum), between 2-4 and 6-7 keV, for both single and double lamp post models. The color convention is the same as in previous plots.}
    \label{fig:lagfreq_comparison}
\end{figure}
\begin{figure}
    \includegraphics[width=\columnwidth, trim={0.65cm 0.0cm 0.6cm 0.5cm},clip]{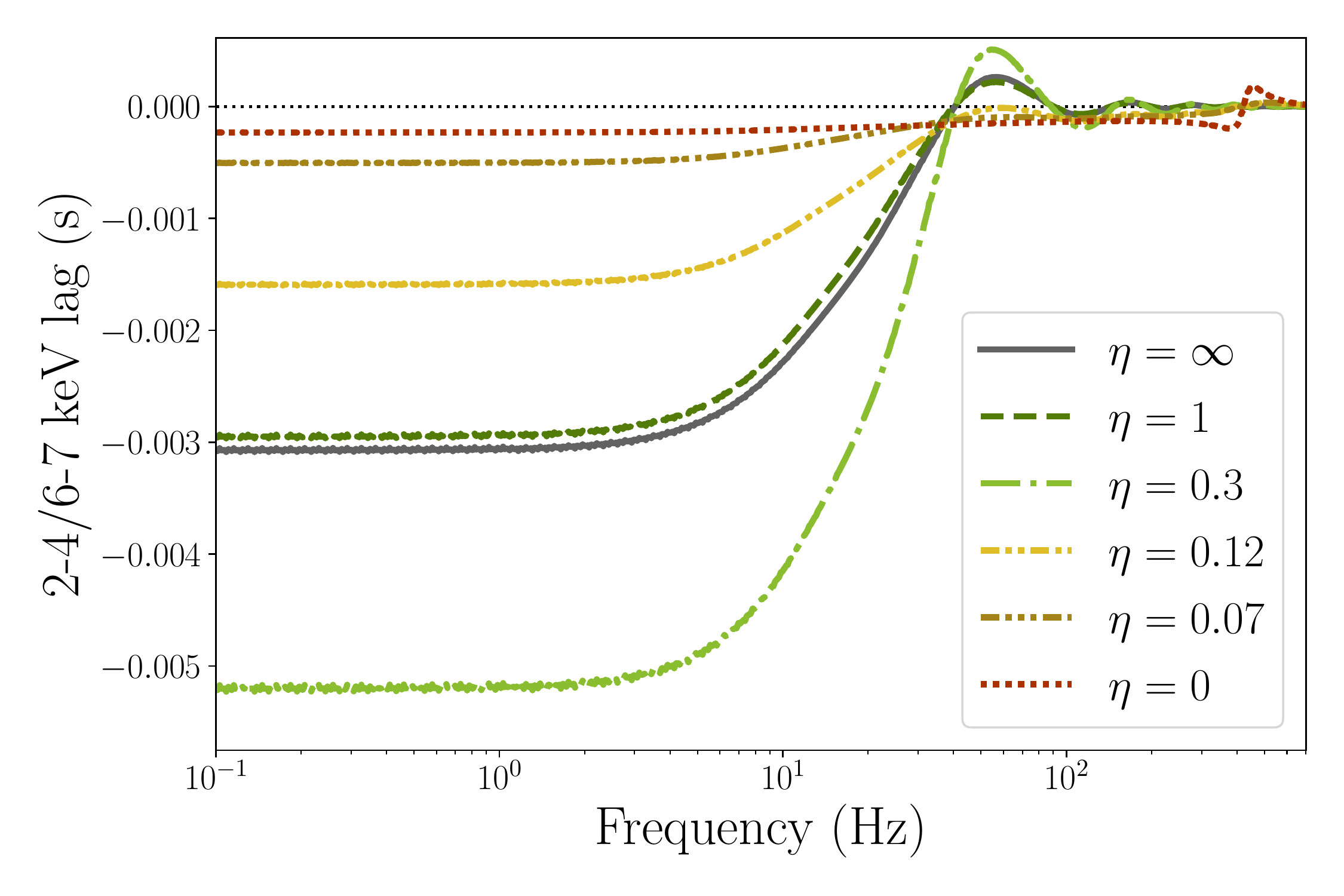}
    \caption{Comparison of lag-frequency spectra for fixed coronal heights, varying $\eta$, and using the same color convention as Fig.\ref{fig:spectra_eta} and \ref{fig:lag_eta}. For $\eta\geq0.01$, phase wraps always occurs at $\approx40$ Hz.}
    \label{fig:lagfreq_eta}
\end{figure}
\begin{table*}
    \centering
    \begin{tabular}{ c | c | c | c }
    Parameter & Model A & Model B & Model C  \\  \hline 
    $h_1\,(\rm{R_g})$  & $35^{+2}_{-3}$ & $16^{+2}_{-1}$ & $21^{+3}_{-2}$ \\
    $h_2\,(\rm{R_g})$  & // & $359^{+53}_{-50}$ & $278^{+49}_{-50}$\\
    $\rm{Incl}\,(\rm{deg})$ & $73^{+2}_{-2}$ & $72^{+1}_{-1}$ & $70^{+1}_{-2}$\\
    $r_{\rm in}\,(\rm{R_g})$ & $1.60^{+0.30}_{-0.36}$ & $\leq 1.26*$ & $\leq 1.18*$ \\
    $\Gamma$  & $1.992^{+0.003}_{-0.004}$ & $2.006^{+0.007}_{-0.007}$ & $2.020^{+0.009}_{-0.008}$ \\
    $\log(\xi_{\rm max})$  & $2.0^{+0.3}_{-0.2}$ & $1.9^{+0.2}_{-0.3}$ & $2.0^{+0.2}_{-0.1}$ \\
    $\log(n_{\rm e})\,(\rm{cm^{-3}})$ & $18.0^{+0.1}_{-0.1}$ & $17.91^{+0.0.07}_{-0.06}$ & $17.97^{+0.03}_{-0.05}$ \\
    $\eta_0$  & // & $0.93^{+0.05}_{-0.05}$ & $0.2^{+0.1}_{-0.1}$  \\
    $\eta$ & // & $0.93$** & $0.96^{+0.07}_{-0.05}$ \\
    $n_{\rm H}\,(\rm{cm}^{-2})$ & $1.3^{+0.1}_{-0.1}\times10^{21}$ & $1.09^{+0.16}_{-0.14}\times10^{21}$ & $1.27^{+0.08}_{-0.11}\times10^{21}$\\
    $1/\mathcal{B}$ & $1.7^{+0.2}_{-0.1}$ & $1.9^{+0.1}_{-0.2}$ & $1.6^{+0.1}_{-0.1}$ \\
    $\rm{Norm}$, \texttt{Reltrans} & $8.67^{+0.09}_{-0.07}\times10^{-2}$ & $8.89^{+0.13}_{-0.12}\times10^{-2}$ & $8.96^{+0.13}_{-0.12}\times10^{-2}$ \\
    $\rm{Norm}$, \texttt{Diskbb} & $7.1^{+0.6}_{-0.6}\times10^{4}$ & $6.2^{+1.0}_{-0.7}\times10^{4}$ & $7.5^{+0.9}_{-0.7}\times10^{4}$\\
    $T_{\rm bb}\,\,\rm{(keV)}$ & $0.294^{+0.004}_{-0.003}$ & $0.296^{+0.004}_{-0.005}$ & $0.287^{+0.005}_{-0.004}$\\  \hline   
    $\chi^{2}/\rm{d.o.f.}$ & $782.31/425=1.84$ & $638.73/423=1.51$ & $616.00/422=1.46$  \\
    \end{tabular}\\
    *: parameter pegged to limit \\
    **: tied parameter
    \caption{Best-fit parameters for all models, excluding the pivoting parameters and phenomenological instrument calibration model.}
    \label{tab:fit_parameters}
\end{table*}
Finally, Fig.\ref{fig:lagfreq_comparison} shows a comparison of the lag-frequency spectra for the same models, again showing exclusively the reverberation signal, and comparing the 2-4 and 6-7 keV bands. In this ideal case, the reverberation lag produced by the double lamp post is somewhat diluted, and its amplitude ($\approx -0.003$ s) is close to but smaller than the large height single lamp post cases. This behavior is unsurprising, as it is consistent with that of the lag-energy spectra. Similarly, the lag amplitude is a strong function of $\eta$, as shown in Fig.\ref{fig:lagfreq_eta}. Once again, the lag increases up to moderate values of $\eta$ up to $\approx -0.005$ s, and then decreases to match the single lamp post model. \textit{These results demonstrate that identical coronal geometries can produce different reverberation lag amplitudes, depending on where the bulk of the variable signal originates from. As a result, the reverberation lag amplitude is not necessarily a good estimator for the light travel time between disk and corona} (and, indirectly, the coronal size). This occurs because coronal regions close to the black hole could also contribute to the total signal, and this dilution leads to lower reverberation lag amplitudes. Instead, the reverberation lag amplitude acts as a proxy for a combination where most of the time-varying signal originates from, independently of the coronal geometry, and how much that signal is diluted for a given geometry. However, both of the plots also show that the frequency at which phase wrapping occurs ($\approx40$ Hz in this case) is only weakly affected by $\eta$, but depends strongly on the coronal size. In the absence of detailed modeling it is this quantity that can be used as a proxy for the coronal height, under the assumption that the lags are originated primarily by reverberation, because unlike the lag amplitude it is unaffected by dilution \citep[e.g.][]{Uttley14,DeMarco21}.

\section{Model application: \MAXI J1820$+$070}
\label{sec:1820}

In this section we apply \texttt{rtransDbl} to a set of \NICER observations of the BHB \MAXI J1820$+$070 first presented in \cite{Wang21}, which also contains the details of the data reduction. We focus on obsID 193--194 (labeled epoch 2 in \citealt{Wang21}), which marked the beginning of the hard-to-soft state transition, and jointly fit the time-averaged spectra with 5 lag-energy spectra. These observations offer a good compromise between large variability, which results in lag-energy spectra with good data quality, long reverberation lags, and a broad iron line, suggesting that the truncation radius of the disk was approaching the ISCO at this point of the outburst.  

As in \cite{Wang21} we ignore the spectral data below 0.5 keV due to uncertainties in the \NICER response. The lag-energy spectra were computed over frequency ranges that do not display any QPO signal; these are 0.1--0.4, 0.5--0.6, 1.1--1.4, 3--4.2, and 4.3--15.6 Hz, respectively. In all lag-energy spectra we take the 0.3-10 keV range as the reference band. For the time-averaged spectrum we include a disk contribution using \texttt{diskbb} As in \cite{Wang21} we re-bin the spectra to include at least 25 counts per bin and to over-sample the instrument resolution by a factor 3. Similarly, we include a systematic error of 1\% below 3 keV as well as the same phenomenological model to handle residual \NICER calibration features present in the time-averaged spectrum: these are near $\approx0.5$ and $\approx2.4$ keV (for which we use the \texttt{edge} model) and at $\approx1.6$ keV (for which we include a gaussian absorption component). We find that our best-fitting parameters are unchanged by the removal of the calibration model, but the fit quality decreases significantly (from $\chi^{2}/\rm{d.o.f.}\approx 1$ to $\approx 1.5$ for the best-fitting model C described below, considering the time-averaged spectrum alone). We assume these features do not contribute to the correlated variability in the lag spectra and therefore do not include them or any systematic error in these data-sets. We account for galactic absorption using the \texttt{tbabs} absorption model. The final model syntax is \texttt{(tbabs*diskbb+reltrans)*gabs*edge*edge} for the time-averaged spectra and just \texttt{reltrans} for the lag-energy spectra\footnote{\texttt{reltrans} includes a \texttt{tbabs} component, so we only need to include it for \texttt{diskbb}}; with this choice, we implicitly assume that the \texttt{diskbb} component is not variable. In every fit we set the black hole spin to its maximal value $a^{*}=0.998$, assume a solar abundance for iron $A_{\rm Fe}=1$, fix the coronal temperature at $kT_{\rm e}=60\,\rm{keV}$, and assume a black hole mass of $10\,M_{\odot}$. We begin with a single lamp post model (using the model flavor \texttt{reltransDCp}), and then apply \texttt{rtransDbl} allowing progressively more model freedom. 

We begin both the single and double lamp post fits by using the \texttt{subplex} fitting algorithm within \texttt{ISIS} version 1.6.51-1 \citep{Houck00} to find parameters close to the best-fit solution, using $\chi^{2}$ statistics. We then explore the parameter space using the \texttt{isis} implementation of the \texttt{emcee} Markov Chain Monte Carlo (MCMC) algorithm \citep{emcee}, starting from the solution found by the least-chi squares fit. We define our best fits from the posterior distributions of the MCMC chains; the details are further described in Appendix A.

Fig.\ref{fig:fits} shows the contribution (for each model) to the total fit statistic from each data-set, normalized by the number of bin. The time-averaged spectrum contains 233 bins, and the lag spectra contain 45 bins each. The best-fitting parameters are listed in tab.\ref{tab:fit_parameters}. The best-fitting pivoting parameters, as well as a comparison of the model and data for each spectrum, are shown in  tab.\ref{tab:pivoting_parameters} and fig.\ref{fig:modelA}--\ref{fig:modelC} in Appendix B. Fig.\ref{fig:fit_comparison} shows the the time-averaged spectrum and high frequency lag energy spectra (4.3--15.6 Hz) for all models, as these are the most challenging spectra to model jointly.

We begin with the standard single lamp post model, which we call Model A. This fit returns a best-fit statistic of $\chi^{2}/\rm{d.o.f}=782.31/425=1.84$, indicating a mediocre fit. As the top panel of Fig.~\ref{fig:fits} shows, the lag-energy spectra between 0.1-0.4~Hz, 3.0-4.2~Hz, and 4.3-15.6~Hz are the ones that are worst fit by the model, while the time-averaged spectrum is reproduced successfully. The main contribution for all the lag spectra comes from the soft lags below 1 keV (as shown in the top right plot of fig.\ref{fig:fit_comparison}). Additionally, some structured residuals are noticeable in the time-averaged spectrum between 6 and 8 keV, suggesting that part of the iron line may not well reproduced (top left plot in fig.\ref{fig:fit_comparison}).
\begin{figure}
    \includegraphics[width=\columnwidth, trim={0.2cm 0.0cm 0.0cm 0.5cm},clip]{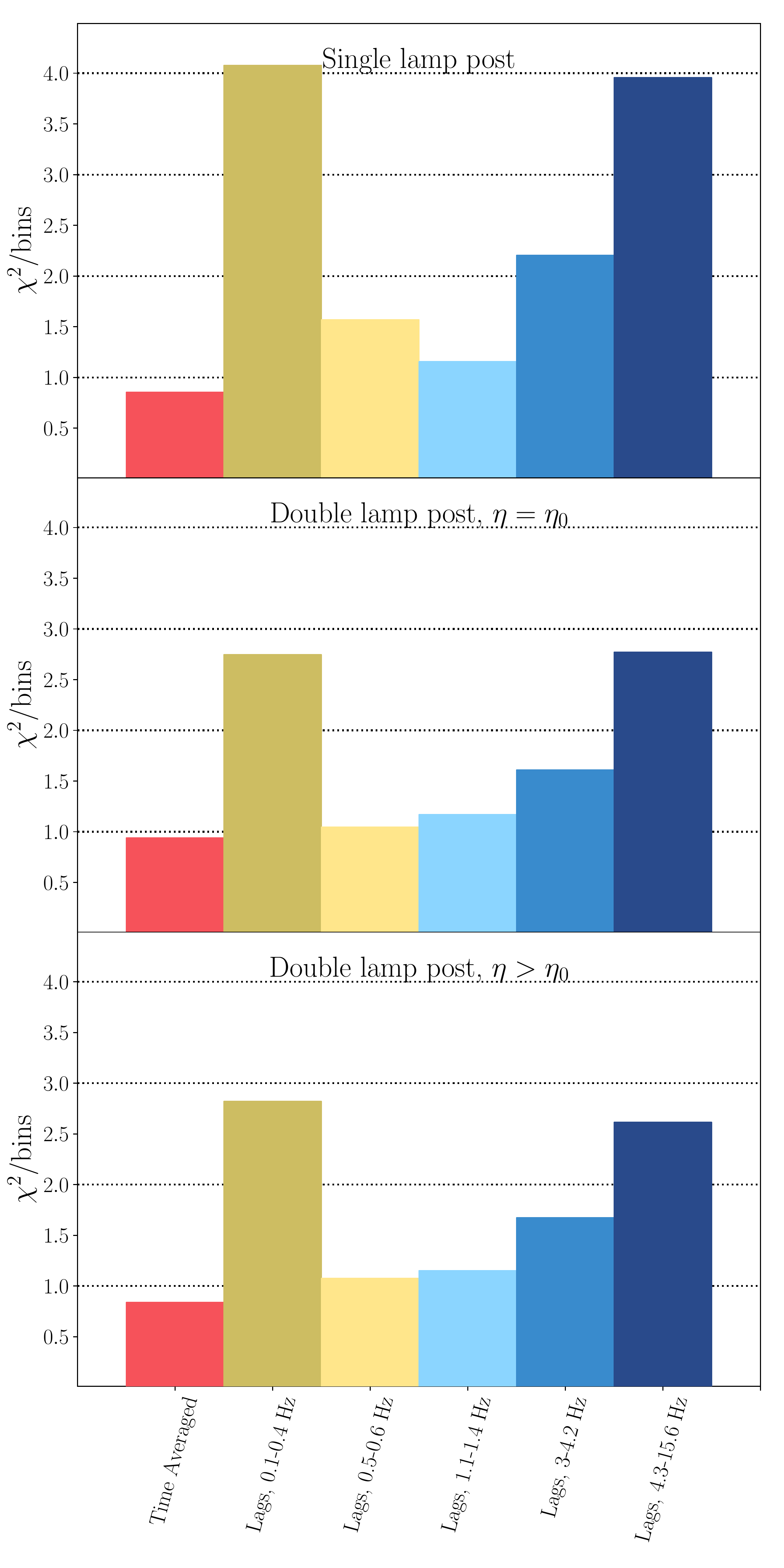}
    \caption{Contribution to the total fit statistic from each spectrum, for models A through C (panels in descending order).}
    \label{fig:fits}
\end{figure}
\begin{figure*}
    \includegraphics[width=\textwidth]{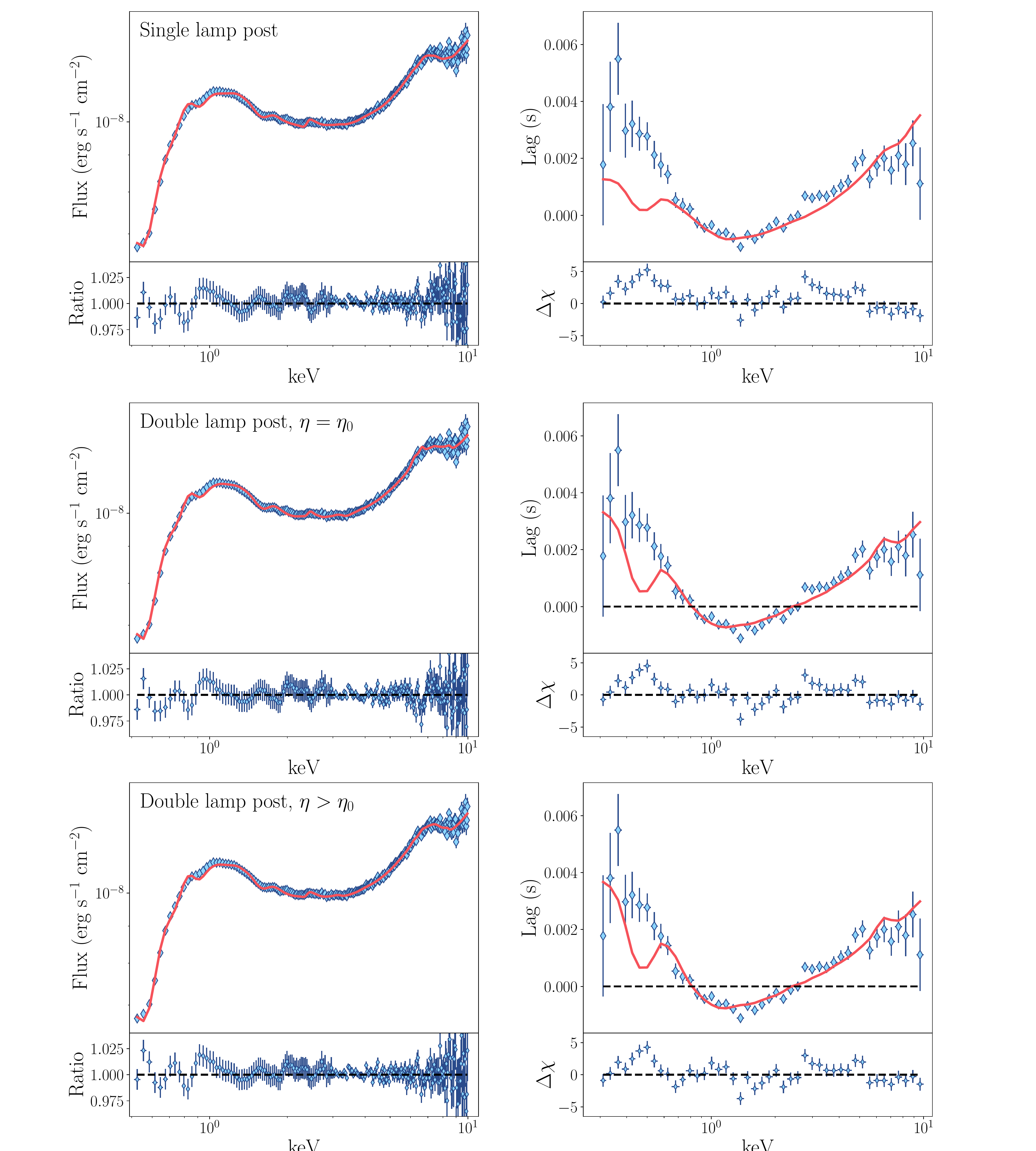}
    \caption{Comparison of the fits to the time-averaged spectra (left column) and 4.3--15.6 lag-energy spectra (right column) , with models A through C (from top to bottom). Model C is the only one that can fit satisfactorily both data-sets.}
    \label{fig:fit_comparison}
\end{figure*}
We then apply the simplest possible instance of the updated double lamp post model, which we call Model B. We begin by setting $\eta=\eta_0$, meaning that the continuum normalization in the comoving frame of each source is identical, as well as tying the pivoting parameters $\phi_{AB,1}=\phi_{AB,2}$ and $\gamma_1=\gamma_2$ in each lag-energy spectrum, meaning that the two sources are assumed to have identical intrinsic variability (meaning that the spectral slope variations depend on flux variations the same way for both sources). With these assumptions we add only two free parameters to the baseline model: the height of the second source $h_2$ and $\eta_0$. Despite the limited number of additional freedom for the model, we find a much more satisfactory fit, with $\chi^{2}/\rm{d.o.f}=638.71/423=1.51$. Fig.\ref{fig:fits} shows that the fit of the lag-energy spectra improves for all Fourier frequencies, particularly in the most problematic frequency ranges, where the fit improves by up to $\Delta \chi^{2}= -59.82$ in the $4.3-15.6\,\rm{Hz}$ frequency range. The improvement at high frequencies in particular suggests that the double lamp-post model captures the reverberation lags (and therefore coronal geometry) more effectively than the single lamp post. On the other hand, compared to model A the time-averaged spectrum fit worsens ($\Delta \chi^{2}= +19.51$); this is mostly caused by residuals around the iron line (middle left plot of fig.\ref{fig:fit_comparison}). The physical reason for this worsening is that fitting the lag-energy spectra requires long reverberation timescales, and therefore a large value of $\eta=0.93$. As a result, the model over-predicts the strength of the narrow component in the time-averaged spectrum.

While Model B improved the overall fit noticeably, the worsening of the statistics for the time-averaged spectrum requires further model complexity. We therefore chose to free $\eta$ from $\eta_0$ in the lag spectra; we refer to this setup as model C. Physically, this choice means that one lamp-post is assumed to be brighter, thus driving a larger time-averaged signal, while the second is assumed to be more variable, thus driving larger lags. At the same time, we keep $\eta_0$ tied across all data-sets, which ensures that the disk ionization state is identical for all spectra. In order to keep the model as simple as possible we tie $\eta$ across all lag-energy spectra, resulting in just one additional free parameter. This choice does not have a strictly physical motivation; rather, in preliminary fits we found that $\eta$ could not be well constrained if we let it vary freely among all lag-energy spectra. The total fit statistic for Model C is $\chi^{2}/\rm{d.o.f}=616.00/422=1.46$, which is a fair improvement over Model B. Unsurprisingly, fig.\ref{fig:fits} shows that the bulk of the improvement in the fit is driven both by the time-averaged ($\Delta\chi^{2}=-3.19$ and $-22.69$ compared to Model A and B, respectively) and 4.3-15.6 Hz lag-energy spectrum ($\Delta\chi^{2}=-60.32$ and $-6.88$ compared to Model A and B, respectively), indicating that Model C captures the both iron line and reverberation lags more effectively than either Models A or B. This improvement is also visible in the bottom left panel of fig.\ref{fig:fit_comparison}). This improvement is mainly driven by the vastly different values of $\eta_0 = 0.2^{+0.1}_{-0.1}$ and $\eta=0.96^{+0.07}_{-0.06}$: the upper corona is dimmer than the lower one, but roughly equally variable. In this way the bottom of the corona mainly drives the time-averaged signal, and the top of the X-ray source is mostly responsible for the variability causing the reverberation signal. 

Finally, the fits show several trends that are largely independent of the model used. First, we find unsurprisingly that the height inferred from Model A is in-between that of the two sources in models B/C. This finding indicates that the vertical extent of the corona is likely playing an important role in the spectral-timing properties of the observation we analyzed. Second, other parameters like the source inclination, photon index, ionization, disk density and boost are broadly consistent throughout the fits, indicating that their values do not impact the properties of the corona inferred from our model. We find a large inclination (in agreement with previous estimates, e.g. \citealt{Torres19}, \citealt{Atri20} and \cite{Espinasse20}, and differing from earlier reflection modeling in \citealt{Buisson19}), a small disk truncation radius, and are able to fit the data without requiring super-Solar iron abundances or large ionization in the outer regions of the disk (although unlike these authors we do not analyze NuSTAR data here). 

\section{Discussion}
\label{sec:discussion}

The main result from our fits is that in order to fit simultaneously the spectra and the time lags, we require a vertically extended corona, where the spectra are dominated by the reflection off the inner accretion disk (irradiated by the lower-height corona), while the time lags are modeled by long light travel times from the upper corona to the disk. In particular, the lags originate from a region farther away from the black hole ($\approx$ hundreds of $R_{\rm g}$ rather than tens in our fits), possibly associated with the ballistic jet launched during the state transition \citep{Wang21,DeMarco21}. This finding is the first test of the scenario proposed initially by \citet{Wang21} and \citet{DeMarco21} for \MAXI J1820, although it appears to be applicable to BHBs as a whole \citep{Wang22}. 

The main caveat to our fits is that our estimates for the height of the second source rely on the assumption that they are the only source of (soft) lags. Our model attempts to match the long reverberation lags with a large source height; the consequence of this choice, however, is that the reflection occurs away from the black hole. As a result, at low ($\leq1\,\rm{keV}$) energies the lag spectra produced by the model are not smooth, but show rather sharp features, unlike the data. This behavior (which is responsible for much of the residuals in the lag spectra, fig.\ref{fig:fits} and \ref{fig:fit_comparison}) is purely a consequence of reflection physics: in our fits, a second lamp post with a large height causes most of the reverberation to originate from the outer disk (Fig.\ref{fig:radial_impulse}), where general relativistic effects are small. This discrepancy with the data could be reduced by introducing additional source of lags in the model. For example, we have assumed that the reflection occurs instantly, while on the other hand the photons take roughly a few $\rm{ms}$ to be re-processed (\citealt{Salvesen22}, Garcia et al., in prep.). Alternatively, it is possible for the balancing of heating and cooling in the corona to produce soft, as well as hard, lags \citep{Uttley18}. Both of these mechanisms produce soft X-ray features similar to those we are associating with light-travel time delays; as a result, including them in our model should result in a lower source height and smoother spectra. We plan to do so in future work. 

If the reverberation lags are indeed originating from the ballistic jet, then the radiative mechanism responsible for the emission of this far region is unlikely to be Comptonization of disk photons. In the case of disk Comptonization (and assuming a small truncation radius), the energy density of disk photons in the co-moving frame of a lamp post at height $h$ scales roughly as $U_{\rm rad}\propto L_{\rm disk}/4\pi h^{2}c$, and as a result the luminosity produced through disk Comptonization will scale as $L_{\rm IC, disk}\propto U_{\rm rad} \propto h^{-2}$: increasing the height by a factor $\approx 10$ implies that, for identical coronal conditions, this luminosity should decrease by a factor $\approx 100$: in other words, if the corona is far from the black hole, then intuitively the number of disk photons reaching it will be greatly reduced. Furthermore, the ballistic jet is moving away from the black hole at relativistic speeds and the seed photons are gravitationally red-shifted, causing the seed photons to be de-boosted; this will further decrease $U_{\rm rad}$, resulting in a further suppression of $L_{\rm IC, disk}$. Instead, it is more likely that the radiative mechanism responsible for the transient jet emission is non-thermal synchrotron emission. If this is the case, then the variability of the transient jet likely originates from particle acceleration through e.g. magnetic reconnection or shocks \citep[e.g.][]{Sironi15,Boettcher19}, which would explain why this outer region appears to be more variable than those near the black hole . 

Recent observations of Cyg X-1 with \IXPE favor a horizontally extended, rather than vertically extended, coronal geometry \citep{Ixpe1}, which would disfavor the model presented here. However, Cyg X-1 is a persistent source and does not behave like a typical transient BHB. In particular, during the \IXPE observations the reflection was relatively weak, unlike BHBs approaching the state transition. Regardless, the main observable \cite{Ixpe1} used to infer the coronal geometry in this source is the X-ray polarization angle being parallel to the VLBI-scale radio jet. On the other hand, \cite{Poutanen23} recently demonstrated that a windy corona outflowing at mildly ($\approx 0.4$c) relativistic speeds can also explain the polarization properties, and such a geometry could be consistent with the vertically extended geometry discussed in this work (although the reflection spectrum would likely be noticeably different due to relativistic beaming, e.g. \citealt{Markoff04,Dauser13}). Observations of AGN meanwhile have produced mixed results: observations of MCG-05-23-16 seem to favor a lamp post or wedge shaped corona over a slab \citep{Marinucci22}, while a lamp post geometry is ruled out in NGC 4151 \citep{Gianolli23}. Finally, observations of the BL Lac Mrk 501 (whose X-ray emission undoubtedly comes from the jet, potentially resulting in a similar polarization signature as a ballistic jet in a BHB) also show that the X-ray polarization angle is parallel to the radio jet \citep{Ixpe2}, which in this case is likely caused by the details of the particle acceleration mechanism operating within the jet.  As such, it is hard to draw strong conclusions or make robust model predictions for such a novel technique. It is likely, however, that if non-thermal synchrotron emission contributes to the observed emission during the state transition, then the polarization degree should increase in these states with respect to the bright hard state. On the other hand, if the X-ray emitting region radiates mainly through Comptonization, then as the state transition happens and the vertical extent of the corona increases one would expect a swing in the X-ray polarization, from parallel to perpendicular to the radio jet.

\section{Conclusion}
\label{sec:conclusion}

In this work we have presented \texttt{rtransDbl}, an updated version of the X-ray spectral timing model \texttt{reltrans}, in which we have extended the lamp post formalism by self-consistently modeling the contribution of a second point source, as a proxy for the vertical extent of the corona. These improvements are a first step towards developing a model that can account for more complex coronal geometries.

We find three main results. First, we demonstrate that the time-averaged spectrum and Fourier-resolved lag-energy spectra can be dominated by different regions of the corona; in particular, it is possible for the bottom of the corona to dominate the time-averaged spectrum, while a large vertical extent of the X-ray emitting region results in long reverberation lags. This behavior offers a possible explanation for the discrepancy in coronal properties inferred from analyzing spectral and timing data alone \citep[e.g.][]{Wang21,Zoghbi21}. 

Second, we find that if the vertical extent of the corona is large enough, then a significant part of the reflected signal is generated in the outer disk and therefore carries weak relativistic effects. This signal therefore provides a possible physical origin for the distant reflection component often identified in the spectra of X-ray binaries \citep[e.g.][]{Garcia13,Garcia15,Degenaar17} . 

Third, by comparing single and double lamp post models we highlight that it is not straightforward to associate a lag amplitude with a coronal size. On the other hand, the Fourier frequency after which the reverberation lags enter the phase wrapping regime can provide a more robust estimate for the source height/vertical extent.

Finally, we compared the single and double lamp post models by using both to model a \NICER observation of the black hole X-ray binary \MAXI J1820+070. We find that in general, the single lamp post model struggles to reproduce the time-averaged spectrum and lag-energy spectra simultaneously, particularly at high Fourier frequencies (which are likely dominated by the reverberation signal). The double lamp post model, on the other hand, provides a much better overall description of the data. In this scenario, we find that the time-averaged spectrum is mainly driven by the bottom of the corona, while the lags are driven by the top source. These proof-of-concept fits represent the first self-consistent test of the phenomenological scenario presented in \cite{Wang21} and \cite{DeMarco21}, and suggest that the inconsistency between from timing and spectral models (e.g. \citealt{Zoghbi21}) might be caused by a vertically extended corona.

\section*{Acknowledgments}

We thank the anonymous referee for insightful comments, which have improved the manuscript. ML, GM, JW, EK and JAG acknowledge support from NASA~ADAP grant 80NSSC17K0515. JW acknowledges support from the NASA~FINNEST Graduate Fellowship, under grant 80NSSC22K1596. AI acknowledges support from the Royal Society. MK acknowledges support by the NWO Spinoza Prize. OK acknowledges funding by the Deutsches Zentrum f{\"u}r Luft-und Raumfahrt, contract 50 QR 2103. This research has made use of \textsl{ISIS} functions (ISISscripts) provided by ECAP/Remeis observatory and MIT (http://www.sternwarte.uni-erlangen.de/isis/). 

\software{Heasoft \citep{Heasoft}), XSPEC  \citep{Arnaud96}, ISIS \citep{isis}, Reltrans \citep{Ingram19}}, Numpy \citep{numpy}, Matplotlib \citep{matplotlib}.

\appendix

\section*{Appendix A: Time-dependent reflection for two lamp posts}
\label{sec:AppendixA}
The form of each transfer function can be derived starting from the linearized spectrum reflected from a single disk patch:
\begin{align}
dR(E,t) \approx dR_0(E_d) + 
\frac{\partial(dR)}{\partial \Gamma_1}\bigg|_0 \delta \Gamma_1(t - \tau_{\rm r_1}) + \frac{\partial(dR)}{\partial \Gamma_2}\bigg|_0 \delta \Gamma_2(t - \tau_{\rm r_2}) + 
\frac{\partial(dR)}{\partial C_1}\bigg|_0 \delta C_1(t - \tau_{r_1}) + \frac{\partial(dR)}{\partial C_2}\bigg|_0 \delta C_2(t - \tau_{r_2}) ,
\label{eq:dRTaylor}
\end{align}
where $\tau_{r_1}= \tau_{\rm s_1 d} + \tau_{\rm do} - \tau_{\rm s_1 o}$ and $\tau_{r_2} = \tau_{\rm s_2 d} + \tau_{\rm do}  - \tau_{\rm s_1 o}$ are the photon arrival times for the reflected photons originated in each lamp post. The second and third terms of eq.\ref{eq:dRTaylor} accounts for changes in the reflection spectrum caused by variations in the continuum photon index(es), and the third for changes in the normalization - this term therefore includes both the reverberation signal, and the lags driven by changes in the ionization discussed in \cite{Mastroserio21}. Let us first evaluate the second and third terms in eq.\ref{eq:dRTaylor}. Note that the \texttt{xillver} rest-frame reflection spectra can only account for a single photon index for the illuminating continuum. On the other hand, if the two sources are pivoting independently, then in general $\Gamma_1(t)\neq\Gamma_2(t)$. As a result, we define:
\begin{equation}
\Gamma(t) = \frac{ C_{1,0} \Gamma_1(t - \tau_{\rm r_1})+ C_{2,0} \Gamma_2(t - \tau_{\rm r_2}) }{ C_{1,0} + C_{2,0} },
\end{equation}
and therefore: 
\begin{equation}
\frac{\partial\mathcal{R}}{\partial\Gamma_1} = \frac{\partial\mathcal{R}}{\partial\Gamma} \frac{\partial\Gamma}{\partial\Gamma_1} = \frac{\partial\mathcal{R}}{\partial\Gamma} \frac{C_{1,0}}{C_0},
\end{equation}
and
\begin{equation}
\frac{\partial\mathcal{R}}{\partial\Gamma_2} = \frac{\partial\mathcal{R}}{\partial\Gamma} \frac{\partial\Gamma}{\partial\Gamma_2} = \frac{\partial\mathcal{R}}{\partial\Gamma} \frac{C_{2,0}}{C_0}.
\end{equation}
Where $C_0 = C_{1,0}+C_{2,0}$. The second and third terms in eq.\ref{eq:dRTaylor} for each source therefore are:
\begin{align}
\frac{\partial(dR)}{\partial \Gamma_i}\bigg|_0 \delta \Gamma_i(t - \tau_{\rm r_i}) =  g_{\rm do}^3 d\alpha d\beta
\bigg\{ \epsilon_i(r) \ln g_{\rm s_i d} C_{i,0}\mathcal{R}  \delta\Gamma_i(t - \tau_{\rm r_i}) 
+ [ C_{1,0} \epsilon_1(r) + C_{2,0} \epsilon_2(r) ]\frac{C_{i,0}}{C_0} \frac{\partial \mathcal{R}}{\partial \Gamma} \delta\Gamma_i(t - \tau_{\rm r_i}) \bigg\}.
\label{eq:Gammapart}
\end{align}
The remaining terms in eq.\ref{eq:dRTaylor} can be evaluated as:
\begin{align}
dR_0(E_d) + \frac{\partial(dR)}{\partial C_1}\bigg|_0 \delta C_1(t^\prime) + \frac{\partial(dR)}{\partial C_2}\bigg|_0 \delta C_2(t^{\prime\prime}) 
\approx \bigg\{ \left[ C_1(t - \tau_{\rm r_1)}) \epsilon_1(r) + C_2(t - \tau_{\rm r_2)}) \epsilon_2(r) \right] \mathcal{R}(E_d) \nonumber \\
+ \left[ C_{1,0}\epsilon_1(r) + C_{2,0}\epsilon_2(r) \right]
\left[ \frac{\partial\xi}{\partial C_1}\bigg|_0 \delta C_1(t - \tau_{\rm r_1)})
+ \frac{\partial\xi}{\partial C_2}\bigg|_0 \delta C_2(t - \tau_{\rm r_2)}) \right]  
\frac{1}{\ln10 \xi_0} \frac{\partial \mathcal{R}}{\partial\log\xi}  \bigg\}  g_{\rm do}^3 d\alpha d\beta,
\label{eq:Awfulpart}
\end{align}
where $\log$ is taken to mean log to the base 10. To differentiate the ionization parameter, we must write it down as a function of $C_1$ and $C_2$, to get
\begin{equation}
\xi(r,t) = \xi_0 \frac{ C_1(t - \tau_{\rm r_1)}) \bar{\epsilon}_1(r) g_{\rm s_1 o}^{\Gamma_0-2} S_1(t - \tau_{\rm r_1)})
+ C_1(t - \tau_{\rm r_2)}) \bar{\epsilon}_2(r) g_{\rm s_2 o}^{\Gamma_0-2} S_2(t - \tau_{\rm r_2)}) } 
{ C_{1,0} \bar{\epsilon}_1(r) g_{\rm s_1 o}^{\Gamma_0-2} + C_{2,0} \bar{\epsilon}_2(r) g_{\rm s_2 o}^{\Gamma_0-2} },
\end{equation}
where
\begin{equation}
S_j(t) \equiv g_{s_j o}^{\delta\Gamma_j(t)} \frac{ \int E^{-\delta\Gamma_j(t)} f(E) dE } { \int f(E) dE },
\end{equation}
and $\bar{\epsilon}_j(r) = \epsilon_j(r) g_{\rm s_j d}^{2-\Gamma}$. The differential of the ionization parameter becomes
\begin{equation}
\frac{\partial \xi}{\partial C_j} = \xi_0 \frac{ \bar{\epsilon}_j(r) g_{\rm s_j o}^{\Gamma_0-2} S_j(t - \tau_{\rm r_1}) } 
{ C_{1,0} \bar{\epsilon}_1(r) g_{\rm s_1 o}^{\Gamma_0-2} + C_{2,0} \bar{\epsilon}_2(r) g_{\rm s_2 o}^{\Gamma_0-2} },
\end{equation}
thus the time averaged value is
\begin{equation}
\frac{\partial \xi}{\partial C_j}\bigg|_0 = \xi_0 \frac{ \bar{\epsilon}_j(r) g_{\rm s_j o}^{\Gamma_0-2} } 
{ C_{1,0} \bar{\epsilon}_1(r) g_{\rm s_1 o}^{\Gamma_0-2} + C_{2,0} \bar{\epsilon}_2(r) g_{\rm s_2 o}^{\Gamma_0-2} }.
\end{equation}
Combining the two equivalent expressions for $\partial \xi / \partial C_j$ gives
\begin{eqnarray}
\left[ C_{1,0}\epsilon_1(r) + C_{2,0}\epsilon_2(r) \right] \left[ \frac{\partial\xi}{\partial C_1}\bigg|_0 \delta C_1(t^\prime)
+ \frac{\partial\xi}{\partial C_2}\bigg|_0 \delta C_2(t^\prime) \right] \frac{1}{\ln10 \xi_0} \frac{\partial \mathcal{R}}{\partial\log\xi} \nonumber \\
 = \frac{ C_{1,0} \epsilon_1(r) + C_{2,0} \epsilon_2(r) } { C_{1,0} \bar{\epsilon}_1(r) g_{\rm s_1 o}^{\Gamma_0-2} + C_{2,0} \bar{\epsilon}_2(r) g_{\rm s_2 o}^{\Gamma_0-2} }
\bigg[ \bar{\epsilon}_1(r) g_{\rm s_1 o}^{\Gamma_0-2} \delta C_1(t - \tau_{\rm r_1)}) + \bar{\epsilon}_2(r) g_{\rm s_2 o}^{\Gamma_0-2} \delta C_2(t - \tau_{\rm r_1)}) \bigg] \frac{1}{\ln10} \frac{\partial \mathcal{R}}{\partial \log\xi}.
\end{eqnarray}
These terms can be re-written more clearly by defining
\begin{equation}
\Theta_j(r) \equiv \bar{\epsilon}_j(r) g_{\rm soj}^{\Gamma_0-2}    
\end{equation}
 and 
 \begin{equation}
 \kappa(r) \equiv \frac{ C_{1,0}\epsilon_1(r) + C_{2,0}\epsilon_2(r) } { C_{1,0} \Theta_1(r) + C_{2,0} \Theta_2(r) },    
 \end{equation}
and as a result we have:
 \begin{align}
dR_0(E_d) + \frac{\partial(dR)}{\partial C_1}\bigg|_0 \delta C_1(t - \tau_{\rm r_1)}) + \frac{\partial(dR)}{\partial C_2}\bigg|_0 \delta C_2(t - \tau_{\rm r_2)}) \nonumber \\
\approx g_{\rm do}^3 d\alpha d\beta \bigg\{
\left[ C_1(t - \tau_{\rm r_1)}) \epsilon_1(r) + C_2(t - \tau_{\rm r_2)}) \epsilon_2(r) \right] \mathcal{R}(E_d) \nonumber \\
+  \kappa(r) \bigg[ \Theta_1(r) \delta C_1(t - \tau_{\rm r_1)}) + \Theta_2(r) \delta C_2(t - \tau_{\rm r_1)}) \bigg]~ \frac{1}{\ln 10} \frac{\partial \mathcal{R}(E_d)}{\partial \log\xi}  \bigg\}.
\label{eq:awfulend}
\end{align}
Here, the first two terms of the right hand side of eq.\ref{eq:awfulend} encode the reverberation signal, and the remaining the lags driven by the varying ionization. Summing eq.\ref{eq:Gammapart} and \ref{eq:awfulend} and taking the Fourier transform returns eq.\ref{eq:fullmodel1} and \ref{eq:fullmodel2}, with the transfer functions given by eq.\ref{eq:w0}--\ref{eq:w3}, in analogy with previous versions of the model. 

\section*{Appendix B: Model parameters}
\label{sec:AppendixB}
\begin{table*}
    \centering
    \begin{tabular}{ c |c | c }
    Parameter name & Description & Value (Sec.2 and 3) \\  \hline 
    $h_1\,(\rm{R_g})$  &  Height of the first lamp post & 2 $\rm{R_g}$\\
    $h_2\,(\rm{R_g})$  &  Height of the second lamp post & 100 $\rm{R_g}$\\
    $a$  &  Black hole spin & 0.998\\
    $\rm{Incl}\,(\rm{deg})$  &  Source viewing angle & 30$^{\circ}$\\
    $r_{\rm in}\,(\rm{R_g})$  &  Disk truncation radius & $\rm{R_{isco}} =1.24 \rm{R_g}$\\
    $r_{\rm out}\,(\rm{R_g})$  &  Outer disk radius & 1000 $\rm{R_g}$\\
    $z$  &  Source redshift & 0\\
    $\Gamma$  &  Continuum photon index & 2\\
    $\log(\xi_{\rm max})$  &  Peak disk ionization value & 3\\
    $A_{\rm fe}$  &  Disk iron abundance & 1\\
    $kT_{\rm e}\,(\rm{keV})$  &  Temperature of the electrons, & 60 $\rm{keV}$\\
    $\log(n_{\rm e})\,\rm{cm}^{-3}$  &  Lowest disk density value & $10^{15}\,\rm{cm}^{-3}$\\
    $\eta_0$  &  Time-averaged normalization ratio & \\
    & $\langle C_{1}\rangle / \langle C_{2}\rangle $ between the two lamp posts, &\\
    & sets continuum cutoff and disk ionization &\\
    $\eta(\nu_{\rm c})$  &  Fourier-frequency dependent & \\
    & normalization ratio $C_{1}(\nu_{\rm c})/C_{2}(\nu_{\rm c})$ &\\
    $n_{\rm H}\,(\rm{cm}^{-2})$  &  Equivalent hydrogen column density & 0 $\rm{cm}^{-2}$\\
    $1/\mathcal{B}$  &  Boost parameter; similar to & 1\\
    & a reflection fraction &\\
    $M_{\rm bh}\,(\rm{M_{\sun}})$  &  Black hole mass & 10 $\rm{M_{\sun}}$\\
    $\nu_{\rm min}\,(\rm{Hz})$  &  Lowest Fourier frequency bound & 1 $\rm{Hz}$\\
    & for frequency averaging & \\
    $\nu_{\rm max}\,(\rm{Hz})$  &  Highest Fourier frequency bound & 10 $\rm{Hz}$\\
    & for frequency averaging &\\
    $\rm{ReIm}$ & Model output format &\\
    $\Phi_{\rm A}(\nu_{\rm c})$  &  Instrument calibration & 0\\
    & phase normalization &\\
    $\Phi_{\rm AB,1}(\nu_{\rm c})$  &  Phase between variations in continuum & 0\\
    & normalization and photon index &\\
    & for the first lamp post &\\
    $\gamma_{\rm 1}(\nu_{\rm c})$  &  Amplitude ratio between spectral & 0\\ 
    & index and normalization variations &\\
    & for the first lamp post &\\
    $\Phi_{\rm AB,2}(\nu_{\rm c})$  &  Phase between variations in continuum & 0\\
    & normalization and photon index &\\
    & for the second lamp post &\\
    $\gamma_{\rm 2}(\nu_{\rm c})$  &  Amplitude ratio between spectral 0 &\\ 
    & index and normalization variations &\\
    & for the second lamp post &\\
    $\alpha(\nu_{\rm c})$  &  Cross spectrum normalization constant; &\\
    & Set to unity for calculating lags &\\
    \end{tabular}
    \caption{Model parameter description, and values used throughout Sec.\ref{sec:model} and \ref{sec:behavior}.}
    \label{tab:model_parameters}
\end{table*}

\begin{table*}
    \centering
    \begin{tabular}{ c | c | c | c }
    Parameter & Model A & Model B & Model C \\  \hline 
    $\Phi_{\rm A}(0.1-0.4\,\rm{Hz})$ & $-1.4^{+0.2}_{-0.1}\times10^{-2}$ &  $-1.1^{+0.1}_{-0.2}\times10^{-2}$ & $-1.1^{+0.2}_{-0.1}\times10^{-2}$ \\
    $\Phi_{\rm AB,1}(0.1-0.4\,\rm{Hz})$& $-5.4^{+0.8}_{-0.3}\times10^{-2}$ &  $-6.3^{+0.5}_{-0.5}\times10^{-2}$ &  $-6.0^{+0.5}_{-0.7}\times10^{-2}$ \\
    $\gamma_{\rm 1}(0.1-0.4\,\rm{Hz})$  & $\geq 0.48^{*}$ & $\geq 0.47^{*}$ &  $\geq 0.46^{*}$  \\
    $\Phi_{\rm AB,2}(0.1-0.4\,\rm{Hz})$ & // & ** & ** \\
    $\gamma_{\rm 2}(0.1-0.4\,\rm{Hz})$ & // & ** & **  \\ 
    $\Phi_{\rm A}(0.5-0.6\,\rm{Hz})$& $-1.6{+0.2}_{-0.3}\times10^{-2}$ & $-1.0^{+0.3}_{-0.3}\times10^{-2}$ & $-1.0^{+0.3}_{-0.3}\times10^{-2}$  \\
    $\Phi_{\rm AB,1}(0.5-0.6\,\rm{Hz})$ & $-6.0^{+0.5}_{-0.5}\times10^{-2}$ &  $-6.4^{+0.7}_{-0.8}\times10^{-2}$ &  $-6.4^{+0.9}_{-1.0}\times10^{-2}$ \\
    $\gamma_{\rm 1}(0.5-0.6\,\rm{Hz})$  & $\geq 0.48^{*}$ & $\geq 0.46^{*}$ &  $\geq 0.46^{*}$ \\
    $\Phi_{\rm A}(1.1-1.4\,\rm{Hz})$ & $-3.2^{+0.3}_{-0.3}\times10^{-2}$ & $-1.8^{+0.3}_{-0.3}\times10^{-2}$ & $-1.7^{+0.3}_{-0.2}\times10^{-2}$\\
    $\Phi_{\rm AB,1}(1.1-1.4\,\rm{Hz})$ & $-0.21^{+0.03}_{-0.02}$ & $-0.48^{+0.10}_{-0.14}$ & $-0.48^{+0.10}_{-0.12}\times10^{-1}$  \\
    $\gamma_{\rm 1}(1.1-1.4\,\rm{Hz})$ & $0.37^{+0.02}_{-0.02}$ &  $0.24^{+0.03}_{-0.04}$ &  $0.23^{+0.04}_{-0.03}$ \\
    $\Phi_{\rm A}(3.0-4.2\,\rm{Hz})$ & $-1.7^{+0.4}_{-0.4}\times10^{-2}$ &  $0^{+0.4}_{-0.3}\times10^{-2}$ &  $-1.0^{+3.7}_{-3.6}\times10^{-3}$  \\
    $\Phi_{\rm AB,1}(3.0-4.2\,\rm{Hz})$ & $-0.13^{+0.02}_{-0.02}$ & $-0.34^{+0.08}_{-0.08}$ &  $-0.33^{+0.07}_{-0.07}$ \\
    $\gamma_{\rm 1}(3.0-4.2\,\rm{Hz})$  & $0.41^{+0.02}_{-0.02}$ &  $0.25^{+0.3}_{-0.3}$ &  $0.25^{+0.03}_{-0.03}$  \\
    $\Phi_{\rm AB,2}(3.0-4.2\,\rm{Hz})$ & // & ** & ** \\
    $\gamma_{\rm 2}(3.0-4.2\,\rm{Hz})$ & // & ** & ** \\ 
    $\Phi_{\rm A}(4.3-15.6\,\rm{Hz})$ &  $0^{+1.6}_{-1.7}\times10^{-3}$ &  $1.4^{+0.5}_{-0.5}\times10^{-2}$ &  $1.5^{+0.3}_{-0.4}\times10^{-2}$  \\
    $\Phi_{\rm AB,1}(4.3-15.6\,\rm{Hz})$ & $-1.1^{+0.3}_{-0.2}\times10^{-1}$ & $-0.30^{+0.16}_{-0.10}$ & $-0.41^{+0.13}_{-0.23}$ \\
    $\gamma_{\rm 1}(4.3-15.6\,\rm{Hz})$  & $0.32^{+0.2}_{-0.2}$ & $0.15^{+0.04}_{-0.04}$ & $0.12^{+0.04}_{-0.03}$ \\
    $\Phi_{\rm AB,2}(4.3-15.6\,\rm{Hz})$ & // & ** & **  \\
    $\gamma_{\rm 2}(4.3-15.6\,\rm{Hz})$ & // & ** & ** \\ \hline     $E_{\rm gabs}\,\,\rm{(keV)}$ \texttt{gabs} & $1.56^{+0.02}_{-0.01}$ & $1.59^{+0.02}_{-0.02}$ & $1.58^{+0.02}_{-0.02}$ \\
    $\sigma\,\rm{(keV)}$ \texttt{gabs} & $\geq 0.09^{*}$ & $\geq 0.09^{*}$ & $\geq 0.08^{*}$ \\ 
    $\rm{Norm}$, \texttt{gabs} & $1.0^{+0.2}_{-0.2}\times10^{-2}$ & $0.9^{+0.1}_{-0.2}\times10^{-2}$ & $0.9^{+0.2}_{-0.1}\times10^{-2}$\\
    $E_{\rm c}\,\,\rm{(keV)}$ \texttt{Edge 1} & $2.38^{+0.02}_{-0.02}$ & $2.41^{+0.03}_{-0.03}$ & $2.38^{+0.03}_{-0.02}$ \\
    $\rm{max}\,\tau$ \texttt{Edge 1} & $-3.8^{+0.2}_{-0.4}\times10^{-2}$ & $-2.9^{+0.5}_{-0.5}\times10^{-2}$ & $-3.1^{+0.5}_{-0.5}\times10^{-2}$ \\
    $E_{\rm c}\,\,\rm{(keV)}$ \texttt{Edge 2} & $0.51^{+0.01}_{-0.01}$ & $0.51^{+0.01}_{-0.01}$ & $0.515^{+0.007}_{-0.008}$ \\
    $\rm{max}\,\tau$ \texttt{Edge 2} & $0.10^{+0.10}_{-0.08}$ & $3.2^{+1.4}_{-1.4}\times10^{-1}$ & $0.19^{+0.09}_{-0.08}$
    \end{tabular}\\
    *: parameter pegged to limit \\
    **: tied parameter
    \caption{Best-fit parameters for the pivoting continuum, for individual lag-energy spectra, \textbf{and for the calibration model for the time-averaged spectrum}.}
    \label{tab:pivoting_parameters}
\end{table*}

Tab.\ref{tab:model_parameters} includes a description of all model parameters, as well as the values used to produce the plots throughout Sec.\ref{sec:model} and \ref{sec:behavior}.

\section*{Appendix C: MCMC setup}
\label{sec:AppendixC}
%TBD update this: 
We found that ``best-guess'' solutions from any least-squares fit are in general not close to the global minimum, regardless of the fitting algorithm or software used (we also fit the data using the Levenberg-Marquardt algorithm in \texttt{Xspec}, finding identical results). These issues remain even when running typical commands to explore the parameter space like \texttt{steppar} or \texttt{error}. On the other hand, the MCMC fit succeeds in shifting the walkers from the initial local minimum, to the true global minimum. The difference in $\chi^{2}$ between the two methods can be very large, up to  $\geq 200$ for all models; similarly, the best-fitting parameters change significantly in value. For example, the height of the lamp post in the single source fit is found to be $h\approx 13\,\rm{R_g}$ by the least-chi squared algorithm (including using \texttt{steppar} to look for a better solution), and $h\approx 35\,\rm{R_g}$ by the MCMC chain.

While the MCMC chain handles the complexity in the parameter space more effectively than a least-chi squared algorithm, recovering the distribution of posteriors near the global minimum is still challenging. In particular, using the default value of the \texttt{emcee} scale parameter $a=2$ (which sets how far the walkers move from one step to the next in the chain, \citealt{emcee}) results in many walkers being stuck in local minima, and as a result the posterior distribution is sampled inefficiently. Due to the computational cost of the model and fit, we tried to limit this behavior as much as possible, as it increases the number of steps required to reach convergence. We found that these issues can be mitigated somewhat by varying the value of $a$. In particular, we found that taking a lower value ($a=1.5$) results in the chain moving most walkers towards the global minimum in about half as many steps, at the cost of leaving a larger number of walkers in local minima. We mitigated this second issue by running an initial chain of 15000 steps, using $a=1.5$. We then run a second chain, using the same number of walkers per free parameters, for 10000 steps but taking $a=2.5$, initialized from the output of the first chain. This choice ended up being more efficient than running a single chain for 30000+ steps for any value of $a$. We conservatively define the posterior distribution from the final 3000 steps, while the initial 22000 steps are discarded as the ``burn-in'' period; we define the best-fitting parameters as the median of the final walker distribution, and the 1-$\sigma$ uncertainty as the range in which 68\% of walkers reside. We always adopt uniform priors within 1\% of the ``best guess'' fit and use 10 walkers per free parameter for all chains.

\section*{Appendix D: Fit information}
\label{sec:AppendixD}
\begin{figure*}[!htbp]
    \includegraphics[width=0.99\textwidth]{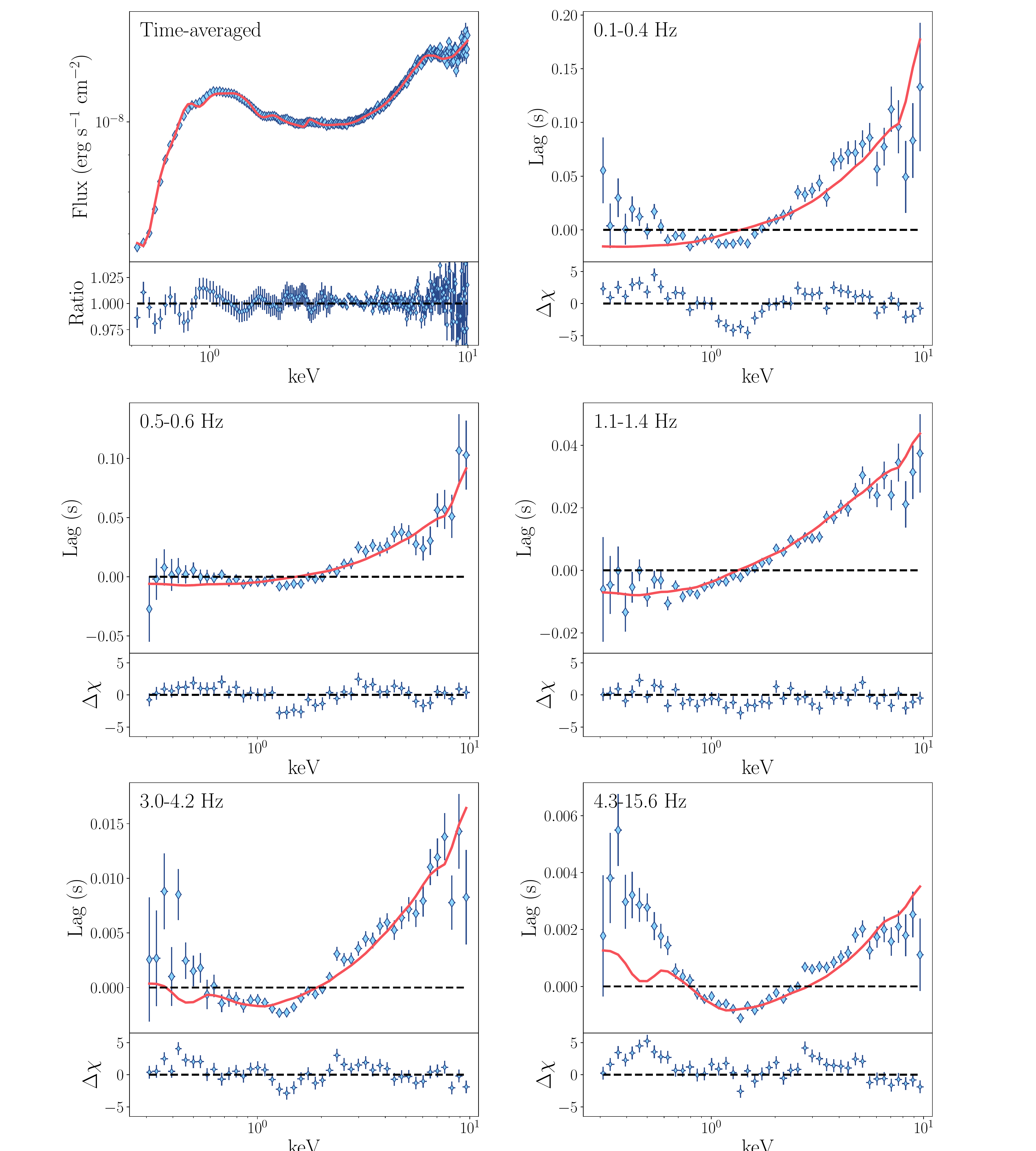}
    \caption{Model A fits and residuals.}
    \label{fig:modelA}
\end{figure*}

\begin{figure*}[!htbp]
    \includegraphics[width=0.99\textwidth]{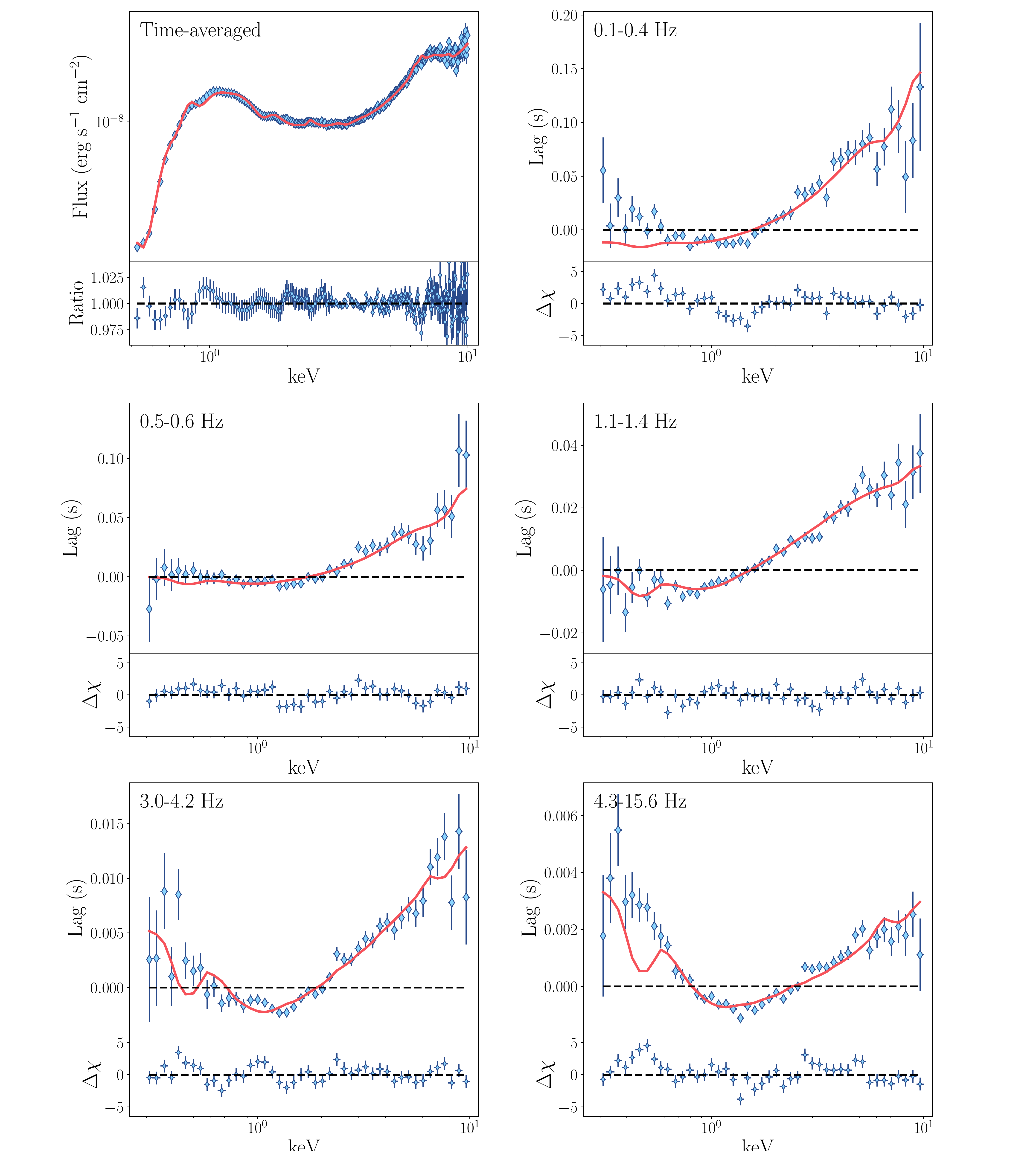}
    \caption{Model B fits and residuals.}
    \label{fig:modelB}
\end{figure*}

\begin{figure*}[!htbp]
    \includegraphics[width=0.99\textwidth]{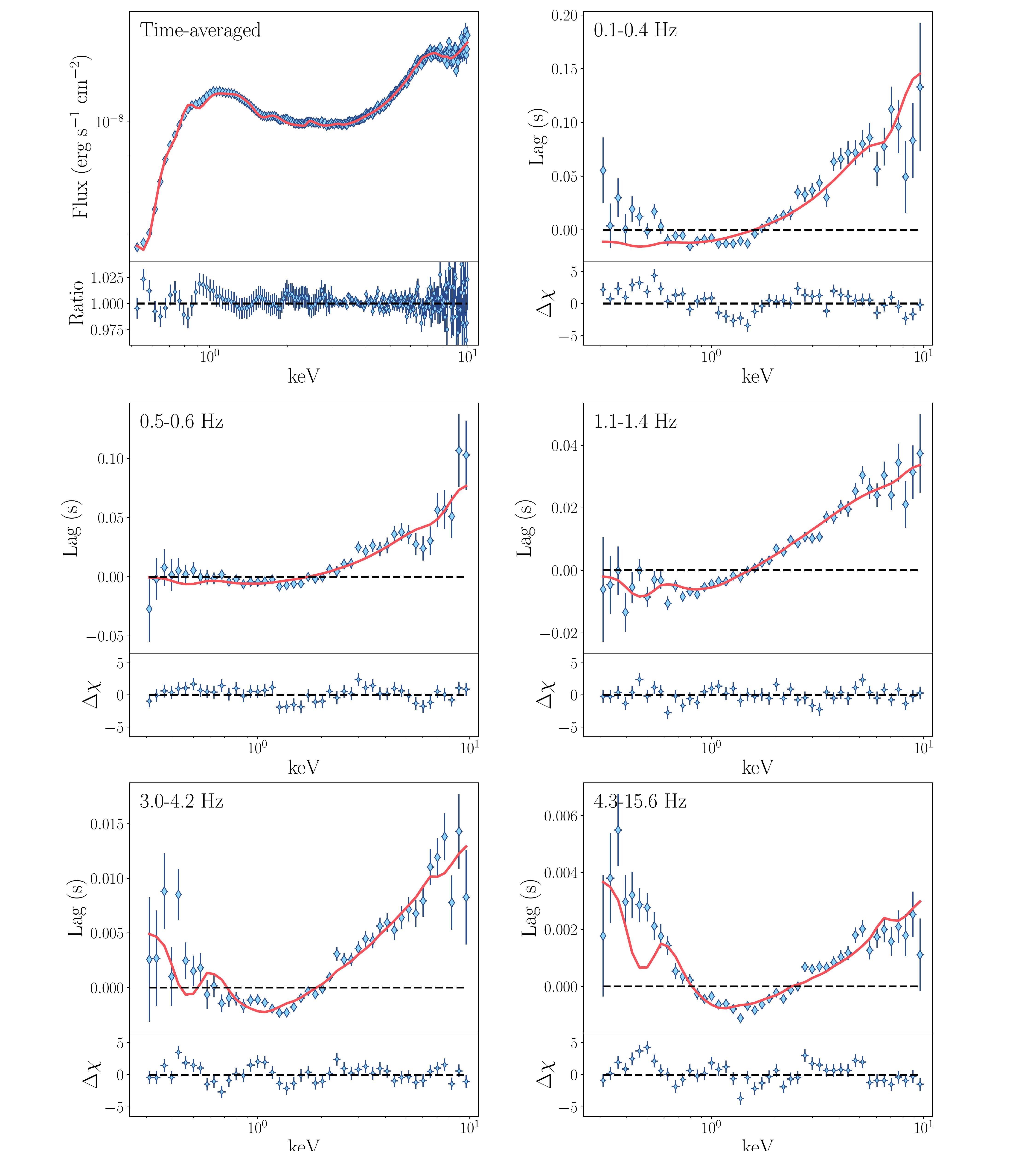}
    \caption{Model C fits and residuals.}
    \label{fig:modelC}
\end{figure*}

Tab.\ref{tab:pivoting_parameters} shows the best fitting parameters for the pivoting power-law(s) for each of the four models. Fig.\ref{fig:modelA}, \ref{fig:modelB} and \ref{fig:modelC} show the fits to the time-averaged spectrum, as well as each lag-energy spectrum, for models A, B and C respectively.

 \clearpage

\bibliographystyle{mnras}
\bibliography{references}
\end{document}